\documentclass[a4paper,11pt]{article}

\usepackage{jheppub} 

\usepackage[T1]{fontenc} 

\title{}

\author[a,b]{Marco Bochicchio}
\author[c]{Samuele P. Muscinelli}
\affiliation[a]{Scuola Normale Superiore (SNS) \\Piazza dei Cavalieri 7, Pisa, I-56100, Italy}
\affiliation[b]{INFN Roma 1\\Piazzale A. Moro 2, Roma, I-00185, Italy}
\affiliation[c]{Dipartimento di Fisica, Universit\`a di Roma `Sapienza'\\Piazzale A. Moro 2, Roma, I-00185, Italy}

\emailAdd{marco.bochicchio@roma1.infn.it}
\abstract{ We point out that perturbation theory
in conjunction with the renormalization group ($RG$) puts a severe constraint on the structure of the large-$N$ non-perturbative glueball propagators in $SU(N)$ pure $YM$, in $QCD$ and
in $\mathcal{N}$ $=1$ $SUSY$ $QCD$ with massless quarks, or in any confining asymptotically-free gauge theory massless in perturbation theory. For the scalar and pseudoscalar glueball propagators
in pure $YM$ and $QCD$ with massless quarks we check in detail the $RG$-improved estimate to the order of the leading and next-to-leading logarithms by means of a remarkable three-loop computation by Chetyrkin et al. We investigate as to whether the aforementioned constraint is satisfied by any of the scalar or pseudoscalar glueball propagators
computed in the framework of the $AdS$ String/ large-$N$ Gauge Theory correspondence and of a recent proposal
based on a Topological Field Theory underlying the large-$N$ limit of $YM$.
We find that none of the proposals for the scalar or the pseudoscalar glueball propagators based on the $AdS$ String/ large-$N$ Gauge Theory correspondence satisfies the constraint, actually as expected, since the gravity side of the correspondence is in fact strongly coupled in the ultraviolet. 
On the contrary, the Topological Field Theory satisfies the constraint that follows by the asymptotic freedom.}

\usepackage{braket}
\usepackage{amsmath}
\usepackage{amssymb}
\usepackage{graphicx}
\usepackage{hyperref}
\usepackage{fancyhdr}
\renewcommand{\Re}{\mathop{\mathrm{Re}}}
\renewcommand{\Im}{\mathop{\mathrm{Im}}}
\newcommand{\Li}{\mathrm{Li}}

\newcommand{\gms}{g_{\overline{MS}}}
\newcommand{\alphams}{\alpha_{\overline{MS}}}
\newcommand{\logp}{\log\frac{p^2}{\mu^2}}
\newcommand{\Lambdams}{\Lambda_{\overline{MS}}}
\newcommand{\Lambdawb}{\Lambda_{\overline{W}}}
\newcommand{\logplw}{\log\frac{p^2}{\Lambda^2_{\overline{W}}}}

\newcommand{\plms}{\frac{p^2}{\Lambdams^2}}
\newcommand{\plw}{\frac{p^2}{\Lambdawb^2}}
\newcommand{\mulms}{\frac{\mu^2}{\Lambdams^2}}

\renewcommand{\Re}{\mathop{\mathrm{Re}}}
\renewcommand{\Im}{\mathop{\mathrm{Im}}}

\title{Ultraviolet asymptotics of glueball propagators}

\begin{document}
\maketitle


\section{Introduction and Conclusions} 

\subsection{Introduction}

In the last years several proposals for the non-perturbative glueball propagators of $QCD$-like confining asymptotically-free gauge theories have been advanced, based
on the $AdS$ String/ large-$N$ Gauge Theory correspondence \cite{Mal} and more recently on a Topological Field Theory ($TFT$) underlying the large-$N$ limit of the pure Yang-Mills ($YM$) theory \cite{boch:quasi_pbs} \cite{MB0} \cite{boch:crit_points} \cite{boch:glueball_prop} \cite{Top}. \par
Above all these proposal aim to elucidate, at least in the large-$N$ limit, the most fundamental feature of the infrared of large-$N$ $QCD$-like confining asymptotically-free gauge theories, i.e. the existence of a mass gap in the pure glue sector, as opposed to the massless spectrum of gluons in perturbation theory. \par
However, these proposals predict a variety of spectra for large-$N$ $QCD$ different among themselves, asymptotically quadratic for large masses \cite{Mal}  \cite{Witten} \cite{Brower1} \cite{PS} \cite{Brower2} or exactly linear \cite{Softwall} \cite{MB0} \cite{boch:crit_points} \cite{boch:glueball_prop} \footnote{Exact linearity in the $TFT$ refers to the joint large-$N$ spectrum of scalar and pseudoscalar glueballs. The $TFT$ in its present formulation does not contain information about higher spin glueballs.}
in the square of the glueball masses and in general do not agree about the qualitative and quantitative details of the low-energy spectrum but for the existence of the mass gap.  \par
In view of the importance of the problem that these proposals aim to answer and in order to discriminate between the various proposals it is worth investigating whether there is any constraint that we know by the fundamental principles of any confining asymptotically-free gauge theory that any supposed answer for the non-perturbative glueball propagators has to satisfy. \par
In fact, we do know with certainty the implications of the asymptotic freedom for the large-momentum asymptotic behavior of any gauge invariant correlation function. \par
In this paper we do not discuss at all the theoretical justification of the various proposal that we examine, leaving it to the original papers.
We limit ourselves to check whether or not the constraint that follows by the asymptotic freedom and by the renormalization group in the ultraviolet ($UV$) is satisfied by any given proposal. Indeed, the importance of this constraint has been pointed out since the early days of large-$N$ $QCD$ \cite{migdal:multicolor}, see also \cite{polyakov:gauge}. \par
In fact the purpose of this paper is threefold. \par

\subsection{Implications of the renormalization group and of the asymptotic freedom}

Firstly, in sect.(2) we point out that perturbation theory in conjunction with the renormalization group ($RG$) severely constraints the asymptotic behavior of glueball propagators in pure $SU(N)$ Yang-Mills,
in $QCD$ and in $\mathcal{N}$ $=1$ $SUSY$ $QCD$ with massless quarks, or in any confining asymptotically-free gauge theory massless to every order of perturbation theory. \par
Indeed, we show in this paper, on the basis of $RG$ estimates, that the most fundamental object involved in the problem of the mass gap \footnote{The lightest glueball is believed to be a scalar
in pure $YM$ and in the 't Hooft large-$N$ limit of $QCD$.}, the scalar ($S$) glueball propagator in any (confining) asymptotically-free gauge theory with no perturbative physical mass scale, up to unphysical contact terms, i.e. distributions supported at coinciding points,
has the following universal, i.e. renormalization-scheme independent, large-momentum asymptotic behavior:
\begin{align}\label{eqn:corr_scalare_inizio}
&\int\langle \frac{\beta(g)}{gN}tr\bigl(\sum_{\alpha\beta}{F_{\alpha\beta}^2}(x)\bigr)\frac{\beta(g)}{gN} tr\bigl(\sum_{\alpha\beta}{F}_{\alpha\beta}^2(0)\bigr)\rangle_{conn}e^{ip\cdot x}d^4x \nonumber\\
= & C_S p^4\Biggl[\frac{1}{\beta_0\log\frac{p^2}{\Lambdams^2}}\Biggl(1-\frac{\beta_1}{\beta_0^2}\frac{\log\log\frac{p^2}{\Lambdams^2}}{\log\frac{p^2}{\Lambdams^2}}\Biggr)+O\biggl(\frac{1}{\log^2\plms}\biggr)\Biggr]
\end{align} 
Analogously for the pseudoscalar ($PS$) propagator:
\begin{align}
&\int\langle \frac{g^2}{N}  tr\bigl(\sum_{\alpha\beta}{F}_{\alpha\beta}\tilde{F}_{\alpha\beta}(x)\bigr) \frac{g^2}{N} tr\bigl(\sum_{\alpha\beta}{F}_{\alpha\beta}\tilde{F}_{\alpha\beta}(0)\bigr)\rangle_{conn}e^{ip\cdot x}d^4x \nonumber\\
=& C_{PS} p^4 \Biggl[\frac{1}{\beta_0\log\frac{p^2}{\Lambdams^2}}\Biggl(1-\frac{\beta_1}{\beta_0^2}\frac{\log\log\frac{p^2}{\Lambdams^2}}{\log\frac{p^2}{\Lambdams^2}}\Biggr)+O\biggl(\frac{1}{\log^2\plms}\biggr)\Biggr]
\end{align}
and for a certain linear combination as well, the anti-seldual ($ASD$) propagator:
\begin{align}\label{eqn:corr_asd}
& \frac{1}{2}\int\langle \frac{g^2}{N}tr\bigl(\sum_{\alpha\beta}F_{\alpha\beta}^{- 2}(x)\bigr)\frac{g^2}{N} tr\bigl(\sum_{\alpha\beta}F_{\alpha\beta}^{- 2}(0)\bigr)\rangle_{conn}e^{ip\cdot x}d^4x \nonumber\\
= &  C_{ADS} p^4 \Biggl[\frac{1}{\beta_0\log\frac{p^2}{\Lambdams^2}}\Biggl(1-\frac{\beta_1}{\beta_0^2}\frac{\log\log\frac{p^2}{\Lambdams^2}}{\log\frac{p^2}{\Lambdams^2}}\Biggr)+O\biggl(\frac{1}{\log^2\plms}\biggr)\Biggr]
\end{align}
where $F_{\alpha\beta}^-=F_{\alpha\beta}-\tilde F_{\alpha\beta}$ and $\tilde F_{\alpha\beta}=\frac{1}{2} \epsilon_{\alpha \beta \gamma \delta} F_{\gamma \delta}$. \par
The explicit dependence on the particular $\Lambdams$ scale in Eq.(\ref{eqn:corr_scalare_inizio})-Eq.(\ref{eqn:corr_asd}) is illusory. A change of scheme affects only the $O\biggl(\frac{1}{\log^2\plms}\biggr)$ terms.
The coincidence of the asymptotic behavior, up to the overall normalization constants that are computed in sect.(3), $C_S, C_{PS},C_{ASD}$, is due to the coincidence of the naive dimension in energy, $4$, and of the one-loop  anomalous dimension, $\gamma(g)=-2 \beta_0 g^2+\cdots$, of these operators deprived of the factors of $\frac{\beta(g)}{g}$ or of $g^2$. Euclidean signature is always understood in this paper unless otherwise specified. \par

\subsection{Perturbative check of the $RG$ estimates}
Secondly, in sect.(3) we check the correctness of our $RG$ estimate on the basis of an explicit very remarkable three-loop computation \footnote{The earlier two-loop computation was performed in \cite{Kataev:1981gr}.}
performed by Chetyrkin et al.\cite{chetyrkin:scalar}
\cite{chetyrkin:pseudoscalar}
in pure $SU(N)$ $YM$ and in $SU(3)$ $QCD$ with $n_f$ massless Dirac fermions in the fundamental representation. 
For example, we show that in pure $SU(N)$ $YM$ Chetyrkin et al. result \cite{chetyrkin:scalar}
\cite{chetyrkin:pseudoscalar} can be rewritten by elementary methods as:
\begin{align}\label{eqn:prologo:ris_sommato}
& \frac{1}{2} \int\langle \frac{g^2}{N} tr\bigl(\sum_{\alpha\beta}{F_{\alpha\beta}^{-2}}(x)\bigr)\frac{g^2}{N} tr\bigl(\sum_{\alpha\beta}{F^{-2}_{\alpha\beta}}(0)\bigr)\rangle_{conn}\,e^{-ip\cdot x}d^4x  \nonumber\\
&=(1-\frac{1}{N^2})\frac{p^4}{2\pi^2\beta_0} \bigl(2\gms^2(\frac{p^2}{\Lambdams^2})-2\gms^2(\frac{\mu^2}{\Lambdams^2}) \nonumber\\
& +\bigl(a+\tilde{a} -\frac{\beta_1}{\beta_0}\bigr)\gms^4(\frac{p^2}{\Lambdams^2})
-\bigl(a+\tilde{a} -\frac{\beta_1}{\beta_0}\bigr)\gms^4(\frac{\mu^2}{\Lambdams^2})\bigr)+O(g^6)
\end{align} 
where $a$ and $\tilde{a}$ are two scheme-dependent constants that are defined in sect.(3.5) and $\gms$ is the 't Hooft coupling constant in the $\overline{MS}$ scheme.
In Eq.(\ref{eqn:prologo:ris_sommato}) the terms that depend on $g(\frac{\mu^2}{\Lambdams^2})$ correspond in the coordinate representation to distributions supported at coincident points (contact terms), and therefore they have no physical meaning.
Remarkably, the correlator without the contact terms does not in fact depend on the arbitrary scale $\mu$ (within $O(g^6)$ accuracy) as it should be.
The running coupling constant $\gms^2(\plms)$ occurs in Eq.(\ref{eqn:prologo:ris_sommato}) with two-loop accuracy and it is given by:
\begin{align}\label{eqn:prologo:g_pert}
&\gms^2(\plms)=\gms^2(\mulms)\bigl(1-\beta_0\gms^2(\mulms)\logp \nonumber\\
&-\beta_1\gms^4(\mulms)\logp +\beta_0^2\gms^4(\mulms) \log^2\frac{p^2}{\mu^2}\bigr) + \cdots
\end{align}
Therefore, the perturbative computation furnishes an expansion of the correlator in powers of $\gms^2(\mu)$ and of logarithms. This expansion has been rearranged by elementary methods in terms of the two-loop running coupling $\gms^2(\plms)$ in Eq.(\ref{eqn:prologo:ris_sommato}). \par
At this point our basic strategy to check the $RG$ estimates of sect.(2) consists in substituting in Eq(\ref{eqn:prologo:ris_sommato}) instead of Eq.(\ref{eqn:prologo:g_pert}) the $RG$-improved expression for $\gms^2(\plms)$ given by:
\begin{equation}\label{eqn:prologo:rg_g}
\gms^2(\plms)=\frac{1}{\beta_0\log\frac{p^2}{\Lambdams^2}}\biggl[1-\frac{\beta_1}{\beta_0^2}\frac{\log\log\frac{p^2}{\Lambdams^2}}{\log\frac{p^2}{\Lambdams^2}}\biggr]
+O\biggl(\frac{\log^2\log\frac{p^2}{\Lambdams^2}}{\log^3\frac{p^2}{\Lambdams^2}}\biggr)
\end{equation}
The $\overline{MS}$ scheme is indeed defined \cite{chetyrkin:schema} in such a way to cancel the term of order of $\frac{1}{\log^2\frac{p^2}{\Lambdams^2}}$ that would occur in Eq.(\ref{eqn:prologo:rg_g}) in other schemes.
By subtracting the unphysical contact terms and by substituting the $RG$-improved two-loop asymptotic expression for $\gms^2(\plms)$ it follows the actual large-momentum scheme-independent asymptotic behavior of Eq.(\ref{eqn:prologo:ris_sommato}):
\begin{align}\label{eqn:prologo_comportamento_pert}
& \frac{1}{2} \int\langle \frac{g^2}{N} tr\bigl(\sum_{\alpha\beta}F_{\alpha\beta}^{- 2}(x)\bigr)\frac{g^2}{N} tr\bigl(\sum_{\alpha\beta}F_{\alpha\beta}^{- 2}(0)\bigr)\rangle_{conn}\,e^{-ip\cdot x}d^4x  \nonumber\\
= & (1-\frac{1}{N^2})\frac{p^4}{\pi^2\beta_0}\Biggl[\frac{1}{\beta_0\log\frac{p^2}{\Lambdams^2}}\Biggl(1-\frac{\beta_1}{\beta_0^2}\frac{\log\log\frac{p^2}{\Lambdams^2}}{\log\frac{p^2}{\Lambdams^2}}\Biggr)+O\biggl(\frac{1}{\log^2\plms}\biggr)\Biggr]
\end{align} 
as opposed to the perturbative behavior that would follow by Eq.(\ref{eqn:prologo:g_pert}).
The asymptotic result in the other cases is checked similarly. \par

\subsection{ $AdS$/Large-$N$ Gauge Theory correspondence and disagreement with the $RG$ estimates}

Thirdly, in this subsection and in the next one, we inquire whether the large-$N$ non-perturbative scalar or pseudoscalar propagators actually computed in the literature agree or disagree with the $RG$ estimate. \par
We find, to the best of our knowledge, that all the scalar propagators presently computed in the literature in the framework of the $AdS$ String/ large-$N$ Gauge Theory correspondence disagree with the universal asymptotic behavior. \par
We should mention that the comparison of the asymptotics of the scalar glueball propagators in the $AdS$ approach with $YM$ or with $QCD$ at the lowest non-trivial order of perturbation theory has been already performed in \cite{forkel} \cite{italiani} \cite{forkel:holograms} \cite{forkel:ads_qcd}, but with somehow different conclusions.
The reasons is that in \cite{forkel}  \cite{italiani} \cite{forkel:holograms} \cite{forkel:ads_qcd} the comparison has been performed only with the one-loop result for the scalar glueball propagator, i.e. only with the first term in Eq.(\ref{eqn:corr_pert_scalar_3l}), that is conformal in the $UV$. No higher order of perturbation theory and no $RG$ improvement has been taken into account in the comparison, as instead we do in this paper. \par
Here we enumerate the models based on the $AdS$/Gauge Theory correspondence for which we could find explicit computations of the scalar glueball propagator in the literature.  \par
In the Hard Wall model (Polchinski-Strassler background \cite{PS} in the so called bottom-up approach):
\[
\int\braket{trF^2(x)trF^2(0)}e^{-i p \cdot x}d^4x \sim  p^4 \biggl[2\frac{K_1(\frac{p}{ \mu})}{I_1(\frac{p}{\mu})}-\logp\biggr]
\]
where $K_1,I_1$ are the modified Bessel functions \cite{forkel}. The asymptotic behavior \cite{forkel} is conformal in the $UV$:
\begin{align*}
&p^4 \biggl[2\frac{K_1(\frac{p}{ \mu})}{I_1(\frac{p}{ \mu})}-\logp\biggr]
 \sim - p^4\biggl[\logp+ O(e^{- 2 \frac{p}{\mu}}) \biggr]
\end{align*}
with $p=\sqrt{p^2}$. Indeed, as recalled in appendix A, in the coordinate representation:
\[
- \int p^4 \logp e^{i p \cdot x}\frac{d^4p}{(2\pi)^4} \sim \frac{1}{x^8}
\]
and, as observed in \cite{forkel} \cite{italiani} \cite{forkel:holograms} \cite{forkel:ads_qcd}, it matches the one-loop large-$N$ $QCD$ result for the perturbative glueball propagator displayed in the first term of Eq.(\ref{eqn:corr_pert_scalar_3l}). Nevertheless, it disagrees by a factor of $(\log p)^2$ with the correct asymptotic behavior in Eq.(\ref{eqn:corr_scalare_inizio}). \par
The Soft Wall model (bottom-up approach) \cite{Softwall} implies the same leading conformal asymptotic behavior \cite{forkel} \cite{italiani} \cite{forkel:holograms} \cite{forkel:ads_qcd} in the $UV$ for the scalar glueball propagator:
\[
\int\braket{trF^2(x)trF^2(0)}e^{-i p \cdot x}d^4x  
 \sim - p^4\biggl[\logp+O(\frac{\mu^2}{p^2})\biggr]
\] 
that therefore disagrees in the $UV$ by the same factor of $(\log p)^2$. \par
A more interesting example of the $AdS$ string / large-$N$ Gauge Theory correspondence from the point of view of first principles applies to the cascading $\mathcal{N}=1$ $SUSY$ $YM$ theory (top-down approach) \cite{KS1} \cite{KS2}, because in this case the correct asymptotically-free $\beta$ function of the cascading theory is exactly reproduced in the supergravity approximation in the Klebanov-Strassler background \cite{KS1} \cite{KS2}.
Nevertheless, the asymptotic behavior of the scalar correlator is \cite{krasnitz:cascading2} \cite{krasnitz:cascading}:
\begin{align*}
\int\braket{trF^2(x)trF^2(0)}e^{-i p \cdot x}d^4x\sim p^4 \log^3 \frac{p^2}{\mu^2}
\end{align*}
that disagrees by a factor of $(\log p)^4$ with the correct asymptotic behavior in Eq.(\ref{eqn:corr_scalare_inizio}). \par

\subsection{Topological Field Theory and agreement with the $RG$ estimates}

Finally, in sect.(4) we prove that in the large-$N$ limit of pure $SU(N)$ $YM$ the $ASD$ glueball propagator computed in \cite{boch:glueball_prop}\cite{boch:crit_points}\cite{MB0} \footnote{We use here a manifestly covariant notation as opposed to the one of the $TFT$ employed in \cite{boch:glueball_prop}\cite{boch:crit_points}\cite{MB0}.}:
\begin{equation}\label{eqn:formula_prologo}
\frac{1}{2} \int\langle \frac{g^2}{N} tr\bigl(\sum_{\alpha\beta}F_{\alpha\beta}^{-2}(x)\bigr)\frac{g^2}{N} tr\bigl(\sum_{\alpha\beta}F_{\alpha\beta}^{-2}(0)\bigr)\rangle_{conn}\,e^{-ip\cdot x}d^4x =
 \frac{1}{\pi^2}\sum_{k=1}^{\infty}\frac{g_k^4\Lambda_{\overline{W}}^6 k^2}{p^2+k\Lambda_{\overline{W}}^2}
\end{equation}
agrees with the universal $RG$ estimate in Eq.(\ref{eqn:prologo_comportamento_pert}).  \par
Since the proposal for the $TFT$ underlying large-$N$ $YM$ is recent and not widely known we add here a few explanations, but for the purposes of this paper the reader can consider Eq(\ref{eqn:formula_prologo}) just as a phenomenological model
factorizing the $ASD$ glueball propagator on a spectrum linear in the masses squared with certain residues. \par 
Yet, to say it in a nutshell, the rationale behind Eq(\ref{eqn:formula_prologo}) is as follows. In \cite{boch:quasi_pbs} \cite{boch:crit_points} \cite{Top} it is shown that there is a $TFT$ trivial \cite{boch:crit_points}  \cite{Top} at $N=\infty$ underlying the large-$N$ limit of $YM$. At $N=\infty$ the $TFT$ is localized on critical points \cite{MB0} \cite{boch:crit_points}. However, at the first non-trivial $\frac{1}{N}$ order the $ASD$ propagator of the $TFT$ arises computing non-trivial fluctuations around the critical points of the $TFT$ \cite{MB0} \cite{boch:glueball_prop}. \par
In Eq.(\ref{eqn:formula_prologo}) $F^-_{\alpha\beta}$ is the anti-selfdual part of the curvature $F_{\alpha\beta}=\partial_\alpha A_\beta-\partial_\beta A_\alpha +i\frac{g}{\sqrt{N}}[A_\alpha,A_\beta]$ with the canonical normalization defined in Eq.(\ref{eqn:F_canonical}), $\Lambda_{\overline{W}}$ is the renormalization group invariant scale in the scheme in which it coincides with the mass gap and $g_k=g(\frac{p^2}{\Lambda^2_{\overline{W}}}=k)$ is the 't Hooft running coupling constant at the scale of the pole (in Minkowski space-time) in the scheme defined in \cite{boch:quasi_pbs}, that is recalled in sect.(4). In fact, the analysis of the $UV$ behavior of Eq.(\ref{eqn:formula_prologo}) has already been performed at the order of the leading logarithm occurring in Eq.(\ref{eqn:corr_asd}) in \cite{boch:glueball_prop}.
Here we go one step further comparing Eq.(\ref{eqn:formula_prologo}) with Eq.(\ref{eqn:corr_asd}) at the order of the next-to-leading logarithm.
Our basic strategy to obtain the large momentum asymptotics of Eq.(\ref{eqn:formula_prologo}) is as follows. We write the RHS of Eq.(\ref{eqn:formula_prologo}) as a sum of physical terms and contact terms according to \cite{boch:glueball_prop}: 
\begin{align}
\frac{1}{\pi^2}\sum_{k=1}^{\infty}\frac{g_k^4\Lambda_{\overline{W}}^6 k^2}{p^2+k\Lambda_{\overline{W}}^2}=
\frac{1}{\pi^2}p^4 \sum_{k=1}^{\infty}\frac{g_k^4\Lambda_{\overline{W}}^2}{p^2+k\Lambda_{\overline{W}}^2} +
 \frac{1}{\pi^2}\sum_{k=1}^{\infty}g_k^4\Lambda^2_{\overline{W}}(k\Lambda^2_{\overline{W}}-p^2)
\end{align}
The first sum contains the physical terms that in Minkowski space-time carry the pole singularities, while the second sum contains the contact terms, that we ignore in the following.
We now consider only the physical terms and to find the leading $UV$ behavior we use the Euler-McLaurin formula according to the technique first introduced by Migdal \cite{migdal:meromorphization} \footnote{We understand that Migdal technique has been known to him for decades.} and employed in \cite{boch:glueball_prop} :
\begin{equation}
\sum_{k=k_1}^{\infty}G_k(p)=
\int_{k_1}^{\infty}G_k(p)dk - \sum_{j=1}^{\infty}\frac{B_j}{j!}\left[\partial_k^{j-1}G_k(p)\right]_{k=k_1}
\end{equation}
where $B_j$ are the Bernoulli numbers.
In our case the terms proportional to the Bernoulli numbers involve negative powers of $p$ and they are therefore suppressed with respect to the first term which behaves as the inverse of a logarithm, so that we ignore them as well. 
We have:
\begin{align}\label{eqn:prologo_hom_intermedio}
\frac{1}{\pi^2}\sum_{k=1}^{\infty}\frac{g_k^4\Lambda_{\overline{W}}^6 k^2}{p^2+k\Lambda_{\overline{W}}^2}\sim 
& \frac{1}{\pi^2}p^4\int_1^{\infty}\frac{g_k^4\Lambda_{\overline{W}}^2}{p^2+k\Lambda_{\overline{W}}^2}dk \nonumber\\
\sim& \frac{1}{\pi^2}p^4\int_1^{\infty}\frac{1}{\beta_0^2\log^2\frac{k}{c}}
\biggl(1-\frac{2\beta_1}{\beta_0^2}\frac{\log\log\frac{k}{c}}{\log\frac{k}{c}}\biggr)
\frac{dk}{k+\frac{p^2}{\Lambda_{\overline{W}}^2}}
\end{align} 
where we have used the $RG$-improved asymptotic behavior for large $k$ of the running coupling constant $g_k$ at the scale of the $k$-th pole, i.e. on shell (in Minkowski space-time):
\begin{equation}
g^2_k\sim \frac{1}{\beta_0\log\frac{k}{c}}\biggl(1-\frac{\beta_1}{\beta_0^2}\frac{\log\log\frac{k}{c}}{\log\frac{k}{c}}\biggr)
\end{equation}
The constant $c$ is related to the scheme that occurs in the non-perturbative calculation \cite{boch:quasi_pbs}\cite{MB0}\cite{boch:crit_points}\cite{boch:glueball_prop}. The actual value of $c$ is not relevant in this paper since we study only the universal asymptotic behavior.
In sect.(4) we compute the universal leading and next-to-leading behavior of the integral in Eq.(\ref{eqn:prologo_hom_intermedio}) and the result is:
\begin{align}
& \frac{1}{\pi^2}p^4\int_1^{\infty}\frac{1}{\beta_0^2\log^2\frac{k}{c}}
\biggl(1-\frac{2\beta_1}{\beta_0^2}\frac{\log\log\frac{k}{c}}{\log\frac{k}{c}}\biggr)
\frac{dk}{k+\frac{p^2}{\Lambda_{\overline{W}}^2}}\nonumber\\
&= \frac{1}{\pi^2\beta_0}p^4 \biggl[\frac{1}{\beta_0\log\bigl(\frac{1}{c}+\frac{p^2}{c\Lambda_{\overline{W}}^2}\bigr)}-\frac{\beta_1}{\beta_0^3}\frac{\log\log\bigl(\frac{1}{c}+\frac{p^2}{c\Lambda^2_{\overline{W}}}\bigr)}{\log^2\bigl(\frac{1}{c}+\frac{p^2}{c\Lambda^2_{\overline{W}}}\bigr)}\biggr]
+O\biggl(\frac{1}{\log^2\frac{p^2}{\Lambda^2_{\overline{W}}}}\biggr)\nonumber\\
&=\frac{1}{\pi^2\beta_0}p^4\biggl[\frac{1}{\beta_0\log\bigl(\frac{p^2}{\Lambda^2_{\overline{W}}}\bigr)}\biggl(1-\frac{\beta_1}{\beta_0^2}\frac{\log\log\bigl(\frac{p^2}{\Lambda^2_{\overline{W}}}\bigr)}{\log^2\bigl(\frac{p^2}{\Lambda^2_{\overline{W}}}\bigr)}\biggr)
\biggr]+O\biggl(\frac{1}{\log^2\frac{p^2}{\Lambda^2_{\overline{W}}}}\biggr)
\end{align} 

\subsection{Conclusions}

The preceding result, for the $ASD$ glueball propagator computed in the $TFT$ underlying large-$N$ pure $YM$, agrees perfectly in the large-$N$ limit with the universal part of the renormalization group improved expression of the perturbative result Eq.(\ref{eqn:prologo_comportamento_pert}). \par
The agreement is due to the conspiracy between the residues of the poles, that are proportional to the fourth power of the coupling constant renormalized on shell times the fourth power of the glueball mass at the pole, and the exact linearity of the joint scalar and pseudoscalar spectrum of the square of the mass of the glueballs in the $ASD$ correlator of the $TFT$. \par
To the best of our knowledge this is the only non-perturbative result for the scalar or pseudoscalar glueball propagator proposed in the literature that agrees with large-$N$ $YM$ perturbation theory and the renormalization group. \par
While this agreement is not by itself a guarantee of correctness of Eq.(\ref{eqn:formula_prologo})  it deserves further investigations, both at theoretical level
and of further checks. \par
Besides, our analysis shows that the $AdS$/Large-$N$ Gauge Theory correspondence in any of its present strong coupling incarnations, the bottom-up or the top-down approach, for which scalar glueball propagators are available in the literature,
does not capture, not even approximatively, the fundamental ultraviolet feature of $YM$ or of $QCD$ or of any large-$N$  confining asymptotically-free gauge theory in the pure glue sector. \par
While this conclusion is certainly known to some experts (see for just one example \cite{KS2}), we think that it is not widely recognized that constructing theories that are conformal in the ultraviolet, as the Hard or the Soft Wall models, or even with the correct beta
function but in the strong coupling phase, as the Klebanov-Strassler supergravity background, is not at all a good approximation for the correct result in the ultraviolet. In this paper, for the first time with leading and next-to-leading logarithmic accuracy, we have computed quantitatively the measure of the disagreement. \par
Finally, given the disagreement between the propagators of the $TFT$ and the propagators of the $AdS$/Large-$N$ Gauge Theory correspondence
in the infrared for the first few lower-mass glueballs, a careful critical analysis of the two approaches at level of numerical lattice data is needed, and also at theoretical level of further constraints arising by the $OPE$ and by the 
low-energy theorems of Shifman-Vainshtein-Zakharov ($SVZ$).

\section{Renormalization group estimates on the universal behavior of correlators}

\subsection{Definitions}

The  $SU(N)$ pure $YM$ theory is defined by the partition function:
\begin{equation}\label{eqn:Z1}
Z=\int \mathcal{D}A\, e^{-\frac{1}{2g_{YM}^2}\int tr F^2(x) d^4x}
\end{equation}
where we use the simplified notation $tr F^2(x)= \sum_{\alpha\beta} tr \bigl(F_{\alpha\beta}^2\bigr)$.
Introducing the 't Hooft coupling constant $g$ \cite{'t hooft:large_n}:
\begin{equation}
g^2=g^2_{YM}N
\end{equation}
the partition function reads:
\begin{equation}
Z=\int \mathcal{D}A\, e^{-\frac{N}{2g^2}\int tr F^2(x) d^4x}
\end{equation} 
According to 't Hooft \cite{'t hooft:large_n} the large-$N$ limit is defined with $g$ fixed when $N\rightarrow \infty$. \par
For the structure of large-$N$ glueball propagators see \cite{migdal:multicolor}
and for reviews of the large-$N$ limit see \cite{polyakov:gauge} and \cite{makeenko:large_n}.
The normalization of the action in Eq.(\ref{eqn:Z1}) corresponds to choosing the gauge field $A_{\alpha}$ in the fundamental representation of the Lie algebra, with generators normalized as:
\begin{equation}
tr\, (t^a t^b)=\frac{1}{2}\delta^{ab}
\end{equation}
In Eq.(\ref{eqn:Z1}) $F_{\alpha\beta}$ is defined by:
\begin{equation}\label{eqn:F_wilsonian}
F_{\alpha\beta}(x)=\partial_\alpha A_\beta-\partial_\beta A_\alpha + i[A_\alpha,A_\beta]
\end{equation}
We refer to the normalization of the action in Eq.(\ref{eqn:Z1}) as the Wilsonian normalization.
Perturbation theory is formulated with the canonical normalization, obtained rescaling the field $A_\alpha$ in Eq.(\ref{eqn:Z1}) by the coupling constant $g_{YM}=\frac{g}{\sqrt{N}}$:
\begin{align}
A_\alpha \rightarrow g_{YM}A^{can}_\alpha
\end{align}
in such a way that in the action the kinetic term becomes independent on $g$:
\begin{equation}
\frac{1}{2}\int tr (F^2(A^{can})) (x) d^4 x
\end{equation}
where:
\begin{align}\label{eqn:F_canonical}
F_{\alpha\beta}= \partial_\beta A^{can}_\alpha - \partial_\alpha A^{can}_\beta +ig_{YM}[A^{can}_\alpha , A^{can}_\beta]
\end{align}
From now on we will simply write $F_{\alpha \beta}$ for the curvature as a function of the canonical field, without displaying the superscript $can$.

\subsection{A short summary of perturbation theory and of the renormalization group}

We recall the relation between bare and renormalized two-point connected correlators of a multiplicatively renormalizable gauge-invariant scalar operator $\mathcal{O}$ of naive dimension in energy $D$:
\begin{equation}
G^{(2)}(p,\mu,g(\mu))=Z_{\mathcal{O}}^{2}(\frac{\Lambda}{\mu}, g(\Lambda)) G_0^{(2)}(p, \Lambda, g(\Lambda))
\end{equation} 
where $G_0^{(2)}$ is the bare connected correlator in momentum space, computed in some regularization scheme with cutoff $\Lambda$, and $\mu$ is the renormalization scale:
\begin{equation}
G_0^{(2)}(p, \Lambda, g(\Lambda))= \int \braket{\mathcal{O}(x)\mathcal{O}(0)}_{conn} e^{i p \cdot x} d^4x  \equiv\braket{\mathcal{O}(p)\mathcal{O}(-p)}_{conn}
\end{equation} 
Since $YM$ or $QCD$ with massless quarks or $\mathcal{N}$ $=1$ $SUSY$ $YM$ with massless quarks is massless to every order of perturbation theory and since $\mathcal{O}$ has naive dimension $D$ we can write:
\begin{equation}
G^{(2)}(p,\mu, g(\mu))=p^{2D-4}G_{DL}^{(2)}(\frac{p}{\mu}, g(\mu))
\end{equation}  
where $G_{DL}^{(2)}$ is a dimensionless function.
The Callan-Symanzik equation for the dimensionless two-point renormalized correlator 
expresses the independence of the bare two-point correlator with respect to the subtraction point $\mu$:
\begin{equation}
\frac{\mathrm{d} G_0^{(2)}}{\mathrm{d}\log\mu}\Big|_{\Lambda,g(\Lambda)}=0
\end{equation}\begin{equation}\label{eqn:RG_eq}
\left(\frac{\partial}{\partial\log\mu}+\beta(g)\frac{\partial}{\partial g(\mu)}+2\gamma_{\mathcal{O}}(g)\right)G_{DL}^{(2)}(\frac{p}{\mu},g(\mu))=0
\end{equation}
where we have defined the beta function with respect to the renormalized coupling $g(\mu)$: 
\begin{equation}
\beta(g)=\frac{\partial g}{\partial\log\mu}\Big|_{\Lambda,g(\Lambda)}
\end{equation}
and the anomalous dimension:
\begin{equation}
\gamma_{\mathcal{O}}(g)= - \frac{\partial\log Z_{\mathcal{O}}}{\partial\log\mu}\Big|_{\Lambda,g(\Lambda)}
\end{equation}
Eq.(\ref{eqn:RG_eq}) can be rewritten taking into account the dependence of $G_{DL}^{(2)}$ on the momentum $p=\sqrt {p^2}$:
\begin{equation}\label{eqn:RG_eq_p}
\left(\frac{\partial}{\partial\log p}-\beta(g)\frac{\partial}{\partial g}-2\gamma_{\mathcal{O}}(g)\right)G_{DL}^{(2)}(\frac{p}{\mu},g(\mu))=0
\end{equation}
The general solution of Eq.(\ref{eqn:RG_eq_p}) is:
\begin{equation}\label{eqn:solu_rg}
G_{DL}^{(2)}(\frac{p}{\mu},g(\mu))=\mathcal{G}(g(\frac{p}{\mu},g(\mu)))\, e^{2\int_{g(\mu)}^{g(p)}\frac{\gamma_{\mathcal{O}}(g)}{\beta(g)}dg} \equiv
\mathcal{G}_{\mathcal{O}}(g(p)) \, Z^2_\mathcal{O}(\frac{p}{\mu},g(\mu))
\end{equation}
The running coupling $g(\frac{p}{\mu},g(\mu))$, that we briefly denote by $g(p)$, solves:
\begin{equation}\label{eqn:eq_rg_flow}
\frac{\partial g(p)}{\partial \log p}=\beta(g(p))
\end{equation}
with the initial condition $g(1,g(\mu))=g(\mu)$. \par 
The multiplicative renormalized factor $Z_\mathcal{O}(\frac{p}{\mu},g(\mu))$ satisfies:
\begin{equation}
\gamma_{\mathcal{O}}(g)= - \frac{\partial\log Z_{\mathcal{O}}}{\partial\log \mu}
\end{equation}
and from now on it is thought as a finite dimensionless function of $g(\mu)$ and $g(p)$ only $Z_\mathcal{O}(g(p),g(\mu))$:
\begin{equation}
Z_{\mathcal{O}}=e^{\int_{g(\mu)}^{g(p)}\frac{\gamma_{\mathcal{O}}(g)}{\beta(g)}dg}
\end{equation}
Eq.(\ref{eqn:solu_rg}) expresses the solution of the $RG$ equation as a product of a $RG$ invariant ($RGI$) function $\mathcal{G}_{\mathcal{O}}$ of $g(p)$ only and of a multiplicative factor $Z_{\mathcal{O}}^2$ that is determined by the anomalous dimension $\gamma_{\mathcal{O}}(g)$ and by the beta function $\beta(g)$. $\mathcal{G}_{\mathcal{O}}$ and $Z_{\mathcal{O}}$ can be computed order by order in renormalized perturbation theory. \par
From Eq.(\ref{eqn:eq_rg_flow}), that represents the coupling constant flow as a function of the momentum, we obtain the well-known behavior of the $RG$-improved 't Hooft running coupling constant with one- and two-loop accuracy, starting from the one- and two-loop perturbative beta function:
\begin{equation}
\beta(g)=-\beta_0 g^3 - \beta_1 g^5 + \cdots
\end{equation}
where for pure $YM$:
\begin{align}
&\beta_0=\frac{11}{3}\frac{1}{(4\pi)^2}\nonumber\\
&\beta_1=\frac{34}{3}\frac{1}{(4\pi)^4}
\end{align}
With two-loop accuracy we get:
\begin{align}\label{eqn:g_2loop}
&\frac{d g}{d\log p}=-\beta_0 g^3 - \beta_1 g^5 \nonumber\\
\Rightarrow & \int_{g(\mu)}^{g(p)}\frac{1}{\beta_0 g^3}(1-\frac{\beta_1}{\beta_0}g^2)dg=-\log\frac{p}{\mu} \nonumber\\
\Rightarrow & \frac{1}{\beta_0}(\frac{1}{2g(\mu)^2}-\frac{1}{2g(p)^2})
-\frac{\beta_1}{\beta_0^2}\log\frac{g(p)}{g(\mu)}=-\log\frac{p}{\mu}\nonumber\\
\Rightarrow & g^2(p)=\frac{g^2(\mu)}{1+2\beta_0 g^2(\mu)\log\frac{p}{\mu}-2\frac{\beta_1}{\beta_0}g^2(\mu)\log\frac{g(p)}{g(\mu)}}\nonumber\\
\sim & \frac{1}{2\beta_0\log\frac{p}{\mu}}
\left(1+\frac{\beta_1}{\beta_0^2}\frac{\log\frac{g(p)}{g(\mu)}}{\log\frac{p}{\mu}}\right) \sim
\frac{1}{2\beta_0\log\frac{p}{\mu}}
\left(1-\frac{\beta_1}{2\beta_0^2}\frac{\log\frac{g^2(\mu)}{g^2(p)}}{\log\frac{p}{\mu}}\right)\nonumber\\
= & \frac{1}{\beta_0\logp}
\left(1-\frac{\beta_1}{\beta_0^2}\frac{\log\logp}{\logp}\right)+O\biggl(\frac{1}{\log^2\frac{p^2}{\mu^2}}\biggr)
\end{align}
This is the well known actual $UV$ asymptotic behavior of the running coupling constant.
The function $\mathcal{G}_{\mathcal{O}}$ in Eq.(\ref{eqn:solu_rg}) is not known from general principles but can be computed in perturbation theory as a function of $g(\mu)$ and then expressed in terms of $g(p)$, since  $\mathcal{G}_{\mathcal{O}}$ is $RGI$. 
Similarly, we can evaluate $Z_{\mathcal{O}}$ using again the one-loop or two-loop perturbative expressions for $\beta(g)$ and $\gamma_{\mathcal{O}}(g)$:
\begin{equation}
\gamma_\mathcal{O}(g)=-\gamma_{0(\mathcal{O})} g^2 -\gamma_{1(\mathcal{O})}g^4 + \cdots
\end{equation}
With one-loop accuracy:
\begin{equation}
Z^2_{\mathcal{O}}\sim\left(\frac{g^2(p)}{g^2(\mu)}\right)^{\frac{\gamma_{0(\mathcal{O})}}{\beta_0}}\sim
\left(\log(\frac{p}{\mu})\right)^{-\frac{\gamma_{0(\mathcal{O})}}{\beta_0}}
\end{equation}
and with two-loop accuracy we have:
\begin{align}\label{eqn:z_pert}
Z^2_{\mathcal{O}} \sim &\left[\frac{1}{2\beta_0\log\frac{p}{\mu}}\left(1-\frac{\beta_1\log\log\frac{p}{\mu}}{2\beta_0^2\log\frac{p}{\mu}}\right)\right]^{\frac{\gamma_{0(\mathcal{O})}}{\beta_0}}
e^{\frac{\gamma_{1(\mathcal{O})}\beta_0-\gamma_{0(\mathcal{O})}\beta_1}{\beta_0^2}\left[\frac{1}{2\beta_0\log\frac{p}{\mu}}\left(1-\frac{\beta_1\log\log\frac{p}{\mu}}{2\beta_0^2\log\frac{p}{\mu}}\right)\right]}\nonumber\\
\sim &\left(\frac{1}{2\beta_0\log\frac{p}{\mu}}\right)^{\frac{\gamma_{0(\mathcal{O})}}{\beta_0}} 
\left(1-\frac{\gamma_{0(\mathcal{O})}\beta_1\log\log\frac{p}{\mu}}{2\beta_0^3\log\frac{p}{\mu}}\right)\nonumber\\
\times &\left\{1+\frac{\gamma_{1(\mathcal{O})}\beta_0-\gamma_{0(\mathcal{O})}\beta_1}{\beta_0^2}\left[\frac{1}{2\beta_0\log\frac{p}{\mu}}
\left(1-\frac{\beta_1\log\log\frac{p}{\mu}}{2\beta_0^2\log\frac{p}{\mu}}\right)\right]\right\}\nonumber\\
\sim &\left(\frac{1}{2\beta_0\log\frac{p}{\mu}}\right)^{\frac{\gamma_{0(\mathcal{O})}}{2\beta_0}}
\Biggl(1-\frac{\gamma_{0(\mathcal{O})}\beta_1\log\log\frac{p}{\mu}}{2\beta_0^3\log\frac{p}{\mu}}+
\frac{\gamma_{1(\mathcal{O})}\beta_0-\gamma_{0(\mathcal{O})}\beta_1}{2\beta_0^3\log\frac{p}{\mu}}\nonumber\\
&-\frac{(\gamma_{0(\mathcal{O})}\gamma_{1(\mathcal{O})}\beta_0\beta_1-(\gamma_{0(\mathcal{O})}\beta_1)^2)\log\log\frac{p}{\mu}}{4\beta_0^6(\log\frac{p}{\mu})^2}
\Biggr) 
\end{align}
In evaluating the last two expressions we have used the two-loop $RG$-improved expression for $g(p)$ given by Eq.(\ref{eqn:g_2loop}).
From the two-loop $RG$-improved expression in Eq.(\ref{eqn:z_pert}) it follows that the leading and next-to-leading logarithms for $Z_{\mathcal{O}}$ are determined only by $\beta_0$, $\beta_1$  and by $\gamma_{0(\mathcal{O})}$, that are in fact universal, i.e. scheme independent. Indeed, the two-loop coefficient of the anomalous dimension $\gamma_{1(\mathcal{O})}$ does not occur in the first $\log\log\frac{p}{\mu}$ term, but only in terms that have a subleading behavior as powers of  logarithms.
Keeping only up to the next-to-leading term in $Z^2_{\mathcal{O}}$, we obtain for the universal logarithmic behavior of the dimensionless two-point correlator:
\begin{equation}\label{eqn:pert_th_next_to_lead}
G_{DL}^{(2)}(\frac{p}{\mu})\sim\left[\left(\frac{1}{2\beta_0\log\frac{p}{\mu}}\right)
\Biggl(1-\frac{\beta_1\log\log\frac{p}{\mu}}{2\beta_0^2\log\frac{p}{\mu}}\Biggr)\right]^{\frac{\gamma_{0(\mathcal{O})}}{\beta_0}}
\mathcal{G}_{\mathcal O}(g(p))
\end{equation}
Thus our aim, in order to get asymptotic estimates, is to determine the one-loop coefficient of the anomalous dimension $\gamma_{0 (\mathcal{O})}$ and the $RGI$ function $\mathcal{G}_{\mathcal O}$ for our operators $\mathcal{O}$.
 
\subsection{Anomalous dimension of $tr F^2$ and of $trF\tilde{F}$}

The operator $\frac{\beta(g)}{g}tr F^2$
is proportional to the conformal anomaly, that is the functional derivative with respect to a conformal rescaling of the metric of the renormalized effective action that must be $RGI$.
Therefore $\frac{\beta(g)}{g}tr F^2$ is $RGI$ as well. 
Hence its anomalous dimension vanishes and, using the notation of the previous section, the form of its correlator, ignoring possible contact terms that will be taken into account in sect.(3), is:
\begin{equation}
G^{(2)}_{\frac{\beta(g)}{g}F^2}(p,\mu,g(\mu))=
p^4\mathcal{G}_{\frac{\beta(g)}{g}F^2}(g(p)) 
\end{equation} 
On the other hand $tr F^2$  is not $RGI$ and therefore its correlator is:
\begin{equation}
G^{(2)}_{F^2}(p,\mu,g(\mu))=
p^4\mathcal{G}_{F^2}(g(p))Z_{F^2}^2(\frac{p}{\mu},g(\mu))
\end{equation}  
Since the relation between $G^{(2)}_{\frac{\beta(g)}{g}F^2}(p,\mu,g(\mu))$ and $G^{(2)}_{F^2}(p,\mu,g(\mu))$ is:
\begin{equation}\label{eqn:relationGg}
G^{(2)}_{\frac{\beta(g)}{g}F^2}(p,\mu,g(\mu))=\left(\frac{\beta(g)}{g}\right)^2 G^{(2)}_{F^2}(p,\mu,g(\mu)) 
\end{equation}
it follows that $\left(\frac{\beta(g)}{g}\right)^2$ and $Z^2(\frac{p}{\mu}, g(\mu))$ must combine in such a way to obtain a function of $g(p)$ only:
\begin{equation}
\left(\frac{\beta(g(\mu))}{g(\mu)}\right)^2\, Z_{F^2}^2(\frac{p}{\mu}, g(\mu))\, \mathcal{G}_{F^2}(g(p)) =
\mathcal{G}_{\frac{\beta(g)}{g}F^2}(g(p)) 
\end{equation}
To find the anomalous dimension $\gamma_{ F^2}$ of $tr F^2$ we exploit once again the property of $\frac{\beta(g)}{g}tr F^2$ being $RGI$.
Its two-point correlator must indeed satisfy the equation:
\begin{equation}
\left(\frac{\partial}{\partial\log p}-\beta(g)\frac{\partial}{\partial g} - 4\right)G^{(2)}_{\frac{\beta(g)}{g}F^2}(p,\mu,g(\mu))=0
\end{equation}
where the last term occurs because we are considering the complete correlator and not its dimensionless part.
Using Eq.(\ref{eqn:relationGg}) we find the anomalous dimension of $tr F^2$:
\begin{align}
&\left(\frac{\partial}{\partial\log p}-\beta(g)\frac{\partial}{\partial g}-4\right)
\biggl[\left(\frac{\beta(g)}{g}\right)^2 G^{(2)}_{F^2}(p,\mu,g(\mu))\biggr]=0 \nonumber\\
\Rightarrow &\biggl[\left(\frac{\beta(g)}{g}\right)^2\frac{\partial}{\partial\log p}
-\left(\frac{\beta(g)}{g}\right)^2\beta(g)\frac{\partial}{\partial g}\nonumber\\
&-2\beta(g)(\frac{\beta(g)}{g})\frac{\partial}{\partial g}\left(\frac{\beta(g)}{g}\right) - 4 \left(\frac{\beta(g)}{g}\right)^2 \biggr]
G^{(2)}_{F^2}(p,\mu,g(\mu))=0 \nonumber\\
\Rightarrow & \biggl[\frac{\partial}{\partial\log p}
-\beta(g)\frac{\partial}{\partial g}
-2g\frac{\partial}{\partial g}\left(\frac{\beta(g)}{g}\right) - 4\biggr]
G^{(2)}_{F^2}(p,\mu,g(\mu))=0
\end{align}
From this equation it follows:
\begin{equation}
\gamma_{F^2}(g)=g \frac{\partial}{\partial g}\left(\frac{\beta(g)}{g}\right)
\end{equation}
With two-loop accuracy this expression reads:
\begin{equation}\label{eqn:dim_anomala_scalar}
\gamma_{tr F^2}(g)=-2\beta_0 g^2-4\beta_1 g^4 +\cdots
\end{equation} 
Keeping only the first term, we can derive the expression for $Z^2(\frac{p}{\mu},g(\mu))$ with one-loop accuracy:
\begin{equation}
Z^2(\frac{p}{\mu},g(\mu))\sim\frac{g^4(p)}{g^4(\mu)}
\end{equation}
Finally, the correlator of $\frac{\beta(g)}{g}trF^2$, with one-loop accuracy, is:
\begin{equation}\label{eqn:corr_l2_rg}
G^{(2)}_{\frac{\beta(g)}{g}F^2}(p,\mu,g(\mu))=
p^4\beta_0^2 g^4(\mu)\,\frac{g^4(p)}{g^4(\mu)} \, \mathcal{G}_{F^2}(g(p))=
p^4\beta_0^2 g^4(p)\, \mathcal{G}_{F^2}(g(p))
\end{equation}
We can repeat similar calculations for the operator $tr F\tilde{F}$ in order to compute its anomalous dimension, using the property of $g^2 tr F\tilde{F}$ being $RGI$.
Indeed $g^2 tr F\tilde{F}$ is the density of the second Chern class or topological charge.
The Callan-Symanzik equation is:
\begin{align}
&\left(\frac{\partial}{\partial\log p}-\beta(g)\frac{\partial}{\partial g} - 4\right)G^{(2)}_{g^2 F\tilde{F}}(p,\mu,g(\mu))\nonumber\\
&=\left(\frac{\partial}{\partial\log p}-\beta(g)\frac{\partial}{\partial g} - 4 \right)\biggl[g^4 G^{(2)}_{F\tilde{F}}(p,\mu,g(\mu))\biggr]=0
\end{align}
from which we obtain the anomalous dimension of $trF\tilde{F}$:
\begin{equation}\label{eqn:dim_anomala_pseudoscalar}
\gamma_{F\tilde{F}}(g)=2\frac{\beta(g)}{g}=-2\beta_0 g^2 -2\beta_1 g^4 +\cdots
\end{equation}
We notice that while the one-loop anomalous dimensions of $trF^2$ and of $tr F\tilde{F}$ coincide, the two-loop anomalous dimensions are different.
This means that the operator $tr {F^-}^2$ has a well defined anomalous dimension only at one loop, in agreement with the fact that it belongs to the large-$N$ one-loop integrable sector of Ferretti-Heise-Zarembo \cite{ferretti:new_struct}.
Therefore, only the universal part of its correlator, that is determined by the one-loop anomalous dimension and by the two-loop $\beta$ function, can be meaningfully compared with the non-perturbative computation in Eq.(\ref{eqn:intro_formula_L2}).

\subsection{Universal behavior of correlators}\thispagestyle{empty}

Knowing the naive dimension $D$ and the anomalous dimension of a (scalar) operator $\mathcal{O}_D$, we can write the asymptotic form for $p>>\mu$ of its correlator obtained by the $RG$ theory.
Indeed, as we recalled in sect.(2.2), assuming multiplicative renormalizability, the $RG$-improved form of the Fourier transform of the correlator is given by:
\begin{equation}\label{eqn:pert_general_behavior}
G^{(2)}(p^2)=\int\braket{\mathcal{O}_D(x)\mathcal{O}_D(0)}_{\mathit{conn}} e^{i p \cdot x} d^4x=
p^{2D-4}\,\mathcal{G}_{\mathcal{O}_D}(g(p)) \, Z^2_{\mathcal{O}_D}(\frac{p}{\mu},g(\mu))
\end{equation}
where the power of $p$ is implied by dimensional analysis, $\mathcal{G}_{\mathcal{O}_D}$ is a dimensionless function that depends only on the running coupling $g(p)$, and $Z^2_{\mathcal{O}_D}$ is the contribution from the anomalous dimension.
But in fact in general the correlator of $\mathcal{O}_D$ is not even multiplicatively renormalizable because of the presence of contact terms. These terms would affect the $UV$ asymptotic behavior, but they are non-physical and therefore they must be subtracted. In fact, they spoil the positivity of the correlator in Euclidean space in the momentum representation, that is required by the Kallen-Lehmann representation (see the comment below Eq.(\ref{eqn:rg_improved_scalar_2l})). \par
In the coordinate representation of the correlator, for $x \neq 0$, contact terms do not occur. Therefore, a strategy to avoid that contact terms interfere with the $RG$ improvement is to pass to the coordinate scheme \cite{chetyrkin:TF}, where the correlator is multiplicatively renormalizable, to compute its $RG$-improved expression, to go back to the momentum representation, and eventually to subtract the contact terms. \par
In the coordinate representation for $x \neq 0$ the solution of the Callan-Symanzik equation reads:
\begin{equation}\label{eqn:pert_general_behavior_x}
G_{\mathcal{O}_D}^{(2)}(x)=\braket{\mathcal{O}_D(x)\mathcal{O}_D(0)}_{\mathit{conn}}=
{\bigl(\frac{1}{x^2}\bigr)}^{D}\,\mathcal{G}_{\mathcal{O}_D}(g(x))\, Z^2_{\mathcal{O}_D}(x \mu ,g(\mu))
\end{equation}
with $x=\sqrt {x^2}$, where we have denoted by $g(x)$ the running coupling in the coordinate scheme \cite{chetyrkin:TF} and by an abuse of notation we have used the same names $\mathcal{G}$ and $Z$ for the $RGI$ function
and renormalization factor in the coordinate and momentum representation. \par
The function $\mathcal{G}(g(p))$ can be guessed at the lowest non-trivial order,
since the correlator must be conformal at the lowest non-trivial order in perturbation theory, that implies $\mathcal{G}(g(x)) \sim const$. Hence:
\begin{equation}
\mathcal{G}(\frac{p}{\mu}) \sim const \,\log\frac{p}{\mu}
\end{equation}
Indeed, in appendix A we show that $ \int p^{2D-4} \log\frac{p}{\mu} e^{i p \cdot x} d^4p = const (\frac{1}{x^2})^{D} $ that is conformal in the coordinate representation. The explicit dependence on $\mu$,
that contradicts $RG$ invariance in the momentum representation, is due to the fact that the correlator in the momentum representation, as opposed to the coordinate representation, is not really multiplicatively renormalizable because (scale dependent) contact terms arise. This can be understood observing that in the coordinate representation for $x \neq 0$  the lowest-order correlator is independent on the scale $\mu$
but it is not an integrable function, in such a way that its Fourier transform needs a regularization, that introduces the arbitrary scale $\mu$. \par
Naively, we can already derive the leading $UV$ asymptotic behavior:
\begin{align}\label{eqn:naive_rg}
&G^{(2)}_{\mathcal{O}_D}(p^2)\sim p^{2D-4}\logp
\biggl(\frac{g^2(p)}{g^2(\mu)}\biggr)^{\frac{\gamma_{0 (\mathcal{O}_D)}}{\beta_0}}\sim 
p^{2D-4}
(g^2(p))^{\frac{\gamma_{0 (\mathcal{O}_D)}}{\beta_0}-1}
\end{align}
where we have used the fact that $g^2(p)\sim\frac{1}{\log(\frac{p}{\mu})}$. It easy to check that for $D=4$ and $\gamma_{0 (\mathcal{O}_D)}=2 \beta_0$ this estimate coincides with Eq.(\ref{eqn:corr_scalare_inizio})-Eq(\ref{eqn:corr_asd}). \par However, this estimate assumes multiplicative renormalizability in the momentum representation and it does not take into account the occurrence of contact terms in the momentum representation of the correlators. \par Nevertheless, in the next section we confirm by direct computation that after subtracting the contact terms the actual behavior of the scalar and of the pseudoscalar correlator agrees with the estimate in Eq.(\ref{eqn:naive_rg}). 

\section{Perturbative check of the universal behavior of correlators}

In this section we obtain the explicit form of the three-loop correlators of $tr F^2$ and of $tr F\tilde{F}$ starting from their imaginary parts that have been computed in \cite{chetyrkin:scalar} \cite{chetyrkin:pseudoscalar} in the $\overline{MS}$ scheme.
The $\overline{MS}$ scheme can be defined as the scheme in which the two-loop $RG$-improved running coupling does not contain
$\frac{1}{\log^2\frac{p^2}{\Lambda_s^2}}$ contributions \cite{chetyrkin:schema}.
More precisely, we consider the equation for the running coupling constant that follows from the two-loop $\beta$ function:
\begin{equation}
\log\frac{p}{\Lambda_s}=\int_{g(\Lambda_s)}^{g(p)}\frac{dg}{\beta(g)}=\frac{1}{2\beta_0 g^2(p)}+\frac{\beta_1}{\beta_0^2}\log\bigl(g(p)\bigr)+ C +\cdots
\end{equation}
where $C$ is an arbitrary integration constant and $\Lambda_s$ is the $RGI$ scale in a generic scheme $s$.
The value of $C$ in the $\overline{MS}$ scheme is chosen in such a way to cancel the $\frac{1}{\log^2\frac{p^2}{\Lambda_s^2}}$ term in the solution:
\begin{align}
&g_s^2(p)=\frac{1}{\beta_0\log\frac{p^2}{\Lambda_s^2}}\biggl[1-\frac{\beta_1}{\beta_0^2}\frac{\log(\beta_0\log\frac{p^2}{\Lambda_s^2})}{\log\frac{p^2}{\Lambda_s^2}}+\frac{C}{\log\frac{p^2}{\Lambda_s^2}}\biggr] \nonumber\\
\Rightarrow &C=\frac{\beta_1}{\beta_0^2}\log(\beta_0)
\end{align}
The result reported in \cite{chetyrkin:scalar}, for $tr F^2$ in the $SU(3)$ $YM$ theory, is:
\begin{align}\label{eqn:im_scalare}
\Im\braket{tr F^2(p)tr F^2(-p)}_{conn}&=
\frac{8}{4\pi}p^4
\biggl\{1+\frac{\alphams(\mu)}{\pi}\biggl[\frac{73}{4}-\frac{11}{2}\log\frac{p^2}{\mu^2}\biggr]\nonumber\\
&+(\frac{\alphams(\mu)}{\pi})^2\biggl[\frac{37631}{96}
-\frac{363}{8}\zeta(2)-\frac{495}{8}\zeta(3)\nonumber\\
&-\frac{2817}{16}\log\frac{p^2}{\mu^2}+\frac{363}{16}\log^2\frac{p^2}{\mu^2}\biggr]\biggr\}
\end{align}
where $\alpha_s=\frac{g^2_{YM}}{4\pi}$ and  $\alpha_{\overline{MS}}$ is $\alpha_s$ in the $\overline{MS}$ scheme.
Firstly, we want to find from Eq.(\ref{eqn:im_scalare}) the result for the $SU(N)$  $YM$ theory and we want to express the result in terms of the 't Hooft coupling in the $\overline{MS}$ scheme $\gms$.
In fact, this operation is quite easy since it is known, and it can be checked in \cite{chetyrkin:pseudoscalar}, that at this order of perturbation theory the rank of the gauge group enters the result only through the Casimir factor $C_A=N$.
Therefore, to obtain the general result it is sufficient to divide by 3 and to multiply by $N$ the coefficient of $\alphams$ and to divide by 9 and to multiply by $N^2$ the coefficient of $\alphams^2$ .
The factors of $N$ and of $N^2$ are then absorbed in the definition of the 't Hooft coupling constant.
We obtain:
\begin{align}\label{eqn:ris_chet}
&\Im\braket{tr F^2(p)tr F^2(-p)}_{conn}\nonumber\\
&=\frac{N^2-1}{4\pi}p^4
\biggl\{1+\gms^2(\mu)\biggl(\frac{73}{3(4\pi)^2}-2\frac{11}{3(4\pi)^2}\log\frac{p^2}{\mu^2}\biggr)\nonumber\\
&+\gms^4(\mu)\biggl[\frac{37631}{54(4\pi)^4}
-\frac{242}{3(4\pi)^4}\zeta(2)-\frac{110}{(4\pi)^4}\zeta(3)\nonumber\\
&-\frac{313}{(4\pi)^4}\log\frac{p^2}{\mu^2}+\frac{121}{3(4\pi)^4}\log^2\frac{p^2}{\mu^2}\biggr]\biggr\}
\end{align}
From Eq.(\ref{eqn:ris_chet}) we derive the complete expression of the correlator,
assuming the correlator in the form:
\begin{align}\label{eqn:ipotesi}
\braket{tr F^2(p)tr F^2(-p)}_{conn}&=
-\frac{N^2-1}{4\pi^2}p^4\log\frac{p^2}{\mu^2}
\biggl[1+\gms^2(\mu)\biggl(f_0-\beta_0\log\frac{p^2}{\mu^2}\biggr)\nonumber\\
&+\gms^4(\mu)\biggl(f_1+f_2\log\frac{p^2}{\mu^2}+f_3\log^2\frac{p^2}{\mu^2}\biggr)\biggr]
\end{align}
We extract the imaginary part of Eq.(\ref{eqn:ipotesi}) that arises from the imaginary part of the logarithm in Minkowski signature, $\log(-\frac{p^2}{\mu^2})=\log\frac{p^2}{\mu^2}-i\pi$. We obtain:
\begin{align}\label{eqn:ris_ipotesi}
&\Im\braket{tr F^2(p)tr F^2(-p)}_{conn}\nonumber\\
&=\frac{(N^2-1)}{4\pi} p^4\biggl[1+f_0g^2(\mu)+(f_1-f_3\pi^2)g^4(\mu)\nonumber\\
&-2\beta_0g^2(\mu)\log\frac{p^2}{\mu^2}+2f_2g^4(\mu)\log\frac{p^2}{\mu^2}+3f_3g^4(\mu)\log^2\frac{p^2}{\mu^2}\biggr]
\end{align} 
Finally, comparing Eq.(\ref{eqn:ris_chet}) with Eq.(\ref{eqn:ris_ipotesi}) we determine the values of the coefficients $f_i$:
\begin{align}
f_0&=\frac{73}{3(4\pi)^2}\nonumber\\
f_1-f_3\pi^2&=(\frac{37631}{54}-\frac{242}{3}\zeta(2)-110\zeta(3))\frac{1}{(4\pi)^4}\nonumber\\
-2\beta_0&=-2\frac{11}{3(4\pi)^2}\nonumber\\
2f_2&=-\frac{313}{(4\pi)^4}\Rightarrow f_2=-\frac{313}{2(4\pi)^4}\nonumber\\
3f_3&=\frac{121}{3(4\pi)^4}\Rightarrow f_3=\frac{121}{9(4\pi)^4}\Rightarrow f_3=\beta_0^2\nonumber\\
\nonumber\\
\Rightarrow f_1&=(\frac{37631}{54}-110\zeta(3))\frac{1}{(4\pi)^4}
\end{align}
Therefore, the correlator is:
\begin{align}\label{eqn:corr_pert_scalar_3l}
\braket{tr F^2(p)tr F^2(-p)}_{conn}=
&-\frac{(N^2-1)}{4\pi^2}p^4\log\frac{p^2}{\mu^2}
\biggl[1+g^2(\mu)\biggl(f_0-\beta_0\log\frac{p^2}{\mu^2}\biggr)\nonumber\\
&+g^4(\mu)\biggl(f_1+f_2\log\frac{p^2}{\mu^2}+\beta_0^2\log^2\frac{p^2}{\mu^2}\biggr)\biggr]
\end{align}
Similarly, the imaginary part of the correlator of $tr F\tilde{F}$, already written in \cite{chetyrkin:pseudoscalar} for the gauge group $SU(N)$, is:
\begin{align}
\Im\braket{tr F\tilde{F}(p)tr F\tilde{F}(-p)}_{conn}=
&\frac{N^2-1}{4\pi}p^4
\biggl\{1+\frac{\alphams(\mu)}{\pi}\biggl[N\biggl(\frac{97}{12}-\frac{11}{6}\log\frac{p^2}{\mu^2}\biggr)\biggr]\nonumber\\
&+(\frac{\alphams(\mu)}{\pi})^2\biggl[N^2\biggl(\frac{51959}{864}-\frac{121}{24}\zeta(2)
-\frac{55}{8}\zeta(3)\nonumber\\
&-\frac{1135}{48}\log\frac{p^2}{\mu^2}+\frac{121}{48}\log^2\frac{p^2}{\mu^2}\biggr)\biggr]\biggr\}
\end{align} 
We obtain:
\begin{align}\label{eqn:corr_pert_pseudoscalar_3l}
\braket{tr F\tilde{F}(p)tr F\tilde{F}(-p)}_{conn}=
&-\frac{(N^2-1)}{4\pi^2}p^4\log\frac{p^2}{\mu^2}
\biggl[1+\gms^2(\mu)\biggl(\tilde{f}_0-\beta_0\log\frac{p^2}{\mu^2}\biggr)\nonumber\\
&+\gms^4(\mu)\biggl(\tilde{f}_1+\tilde{f}_2\log\frac{p^2}{\mu^2}+\beta_0^2\log^2\frac{p^2}{\mu^2}\biggr)\biggr]
\end{align}
where:
\begin{align}
\tilde{f}_0&=\frac{97}{3(4\pi)^2}\nonumber\\
\tilde{f}_1&=(\frac{51959}{54}-110\zeta(3))\frac{1}{(4\pi)^4}\nonumber\\
-2\beta_0&=-2\frac{11}{3(4\pi)^2}\nonumber\\
2\tilde{f}_2&=-\frac{1135}{3(4\pi)^4}\Rightarrow \tilde{f}_2=-\frac{1135}{6(4\pi)^4}\nonumber\\
\end{align} 

\subsection{Correlator of $\frac{\beta(g)}{gN}tr F^2$ in $SU(N)$ $YM$ (two loops)}

We now determine the $UV$ asymptotic behavior for the correlators by employing their $RG$-improved expression.
Firstly, we recall that in every generic scheme labelled by $a$ the relation between the coupling constant at two different scales is, with  one-loop accuracy:
\begin{align}\label{eqn:scale_relation}
&\frac{1}{g_{a}^2(\mu)}=\frac{1}{g_{a}^2(p)}-\beta_0\logp 
\end{align}
This relation is necessary to express the correlators in their $RG$-improved form.
As a starting simplified example we consider the two-loop expression of the correlator of $\beta_0 \frac{g^2}{N}trF^2$ (for the moment we skip overall positive numerical factors in the normalization of the correlator):
\begin{align}\label{eqn:2l_perturbative_corr}
&\braket{\beta_0 \frac{g^2}{N}tr F^2(p) \beta_0 \frac{g^2}{N}tr F^2(-p)}_{conn}\nonumber\\
&\sim -\beta_0^2 p^4 \gms^4(\mulms)\log\frac{p^2}{\mu^2}
\biggl[1+\gms^2(\mulms)\biggl(f_0-\beta_0\log\frac{p^2}{\mu^2}\biggr)\biggr]
\end{align}
This expression is renormalization group invariant with one-loop accuracy, since the factor $(\frac{\beta(g)}{g})^2$  is $\beta_0^2 g^4$ if we employ the one-loop $\beta$ function. 
The finite term $f_0 \gms^2(\mu)$ can be absorbed in a change of scheme.
Indeed, defining:
\begin{equation}
g^2_{a}(\mu)=\gms^2(\mu)(1+a\gms^2(\mu))
\end{equation}
it follows:
\begin{align}\label{eqn:cambio_schema}
g^4_{a}(\mu)&=\gms^4(1+2a\gms^2(\mu)+a^2\gms^4(\mu))+ O(g^{10})\nonumber\\
\gms^2(\mu)&=g^2_{a}(\mu)(1-a\gms^2(\mu)+a^2\gms^4(\mu))+O(g^8)\nonumber\\
&=g^2_{a}(\mu)(1-a g_{a}^2(\mu)+2a^2g^4_{a}(\mu))+O(g^8)\nonumber\\
\gms^4(\mu)&=g^4_{a}(\mu)(1-2a g^2_{a}(\mu)+5a^2g^4_{a}(\mu))+O(g^{10})
\end{align}
We obtain for the correlator:
\begin{align}
&\braket{\beta_0 \frac{g^2}{N}tr F^2(p) \beta_0 \frac{g^2}{N}tr F^2(-p)}_{conn}\nonumber\\
& \sim -\beta_0^2 p^4\log\frac{p^2}{\mu^2}g^4_{a}(\mu)
\biggl[1+(f_0-2a)g^2_{a}(\mu)-\beta_0g^2_{a}(\mu)\log\frac{p^2}{\mu^2}+O(g^4\logp)\biggr]
\end{align}
To cancel the finite term it is sufficient to put:
\begin{equation}
a=\frac{f_0}{2}
\end{equation}
Hence we obtain:
\begin{align}
&\braket{\beta_0 \frac{g^2}{N}tr F^2(p) \beta_0 \frac{g^2}{N}tr F^2(-p)}_{conn}\nonumber\\
&\sim-\beta_0^2 p^4\log\frac{p^2}{\mu^2}g^4_{a}(\mu)\biggl[1-\beta_0g^2_{a}(\mu)\log\frac{p^2}{\mu^2}+O(g^4\logp)\biggr]
\end{align} 
At this order the term in square brackets is precisely the renormalization factor necessary to renormalize two powers of $g_{a}(\mu)$.
We obtain:
\begin{align}
&\braket{\beta_0 \frac{g^2}{N}tr F^2(p) \beta_0 \frac{g^2}{N}tr F^2(-p)}_{conn}\nonumber\\
&\sim-\beta_0^2 p^4g^2_{a}(\mu)g^2_{a}(p)\log\frac{p^2}{\mu^2}\bigl(1+O(g^4\logp)\bigr)
\end{align} 
From  Eq.(\ref{eqn:scale_relation}) we express the logarithm in terms of the coupling constant:
\begin{equation}
\beta_0\log\frac{p^2}{\mu^2}=\frac{1}{g^2_{a}(p)}-\frac{1}{g^2_{a}(\mu)}
\end{equation}
The correlator becomes:
\begin{align}\label{eqn:rg_improved_scalar_2l}
&\braket{\beta_0 \frac{g^2}{N}tr F^2(p) \beta_0 \frac{g^2}{N}tr F^2(-p)}_{conn}\nonumber\\
&\sim-\beta_0 p^4g^2_{a}(\mu)g^2_{a}(p)\bigl(\frac{1}{g^2_{a}(p)}-\frac{1}{g^2_{a}(\mu)}\bigr)\bigl(1+O(g^4\logp)\bigr)\nonumber\\
&=\beta_0 p^4\bigl(g_{a}^2(p)-g_{a}^2(\mu)\bigr)\bigl(1+O(g^4\logp)\bigr)
\end{align}
The second term in the last line is in fact a contact term that has no physical meaning, therefore it may depend on the arbitrary scale $\mu$ since it must be subtracted anyway.
The physical term is positive, despite the correlator that we started with was negative. 
This is an important feature, since a negative physical term would have been in contrast with the Kallen-Lehmann representation, that requires a positive spectral function.
\newline

\subsection{Correlator of $\frac{\beta(g)}{gN}tr F^2$ in $SU(N)$ $YM$ (three loops)}

We now consider the three-loop result Eq.(\ref{eqn:corr_pert_scalar_3l}), this time including also the correct normalization factors:
\begin{align}
&\braket{\frac{g^2}{N} tr F^2(p)\frac{g^2}{N}tr F^2(-p)}_{conn}\nonumber\\
&=-\bigl(1-\frac{1}{N^2}\bigr)\gms^4(\mu)\frac{p^4}{4\pi^2}\log\frac{p^2}{\mu^2}
\biggl[1+\gms^2(\mu)\biggl(f_0-\beta_0\log\frac{p^2}{\mu^2}\biggr)\nonumber\\
&+\gms^4(\mu)\biggl(f_1+f_2\log\frac{p^2}{\mu^2}+\beta_0^2\log^2\frac{p^2}{\mu^2}\biggr)\biggr]
\end{align}
This correlator is not supposed to be $RGI$, because the factor of $(\frac{\beta(g)}{g})^2=g^4\bigl(1+\frac{\beta_1}{\beta_0}g^2\bigr)^2$ is missing.
We can eliminate the finite terms in the correlator by a redefinition of the scheme:
\begin{align}\label{eqn:cambio_schema2}
g^2_{ab}(\mu)&=  \gms^2(\mu) \bigl(1+a\gms^2(\mu)+b\gms^4(\mu)\bigr)\nonumber\\
\Rightarrow \gms^4(\mu)&=g^4_{ab}(\mu)(1-2a g^2_{ab}(\mu)+(2b+5a^2)g^4_{ab}(\mu))+O(g^{10})
\end{align}
Substituting we obtain:
\begin{align}
&\braket{\frac{g^2}{N} tr F^2(p)\frac{g^2}{N}tr F^2(-p)}_{conn}\nonumber\\
&=-\bigl(1-\frac{1}{N^2}\bigr) g^4_{ab}(\mu)\bigl(1-2a g^2_{ab}(\mu)+(2b+5a^2)g^4_{ab}(\mu)\bigr)
\frac{p^4}{4\pi^2}\log\frac{p^2}{\mu^2}\quad \nonumber\\
\times\quad&\biggl[1+f_0 g^2_{ab}(\mu)(1-ag^2_{ab}(\mu))
-\beta_0 g^2_{ab}(\mu)(1-ag^2_{ab}(\mu))\log\frac{p^2}{\mu^2}+f_1 g_{ab}^4(\mu)\nonumber\\
&+ f_2 g_{ab}^4(\mu)\logp +\beta_0^2 g_{ab}^4(\mu)\log^2\frac{p^2}{\mu^2}\biggr]\nonumber\\
&=-\bigl(1-\frac{1}{N^2}\bigr) g^4_{ab}(\mu)
\frac{p^4}{4\pi^2}\log\frac{p^2}{\mu^2}\quad \nonumber\\
\times\quad&\biggl[1+(f_0-2a) g^2_{ab}(\mu)
-\beta_0g^2_{ab}(\mu)\log\frac{p^2}{\mu^2}+(f_1+5a^2+2b-af_0) g_{ab}^4(\mu)\nonumber\\
&+(f_2+3\beta_0 a) g_{ab}^4(\mu)\logp +\beta_0^2 g_{ab}^4(\mu)\log^2\frac{p^2}{\mu^2}\biggr]
\end{align} 
We eliminate the two finite terms choosing:
\begin{align}
&a=\frac{f_0}{2}\nonumber\\
&f_1+5(\frac{f_0}{2})^2+2b-\frac{f_0^2}{2}=0 \nonumber\\
\Rightarrow & b=\frac{3}{8}f_0^2-\frac{f_1}{2}
\end{align}  
With this choice of $a$ the coefficient of the $g^4\logp$ term becomes:
\begin{equation}
f_2+3\beta_0 a = f_2+\frac{3}{2}f_0\beta_0=-\frac{68}{3(4\pi)^4}=-2\beta_1
\end{equation}
Therefore, the correlator reads:
\begin{align}\label{eqn:corr_scalare_intermedio}
&\braket{\frac{g^2}{N} tr F^2(p)\frac{g^2}{N}tr F^2(-p)}_{conn}=
-\bigl(1-\frac{1}{N^2}\bigr) g^4_{ab}(\mu)
\frac{p^4}{4\pi^2}\log\frac{p^2}{\mu^2}\quad \nonumber\\
\times\quad&\biggl[1-\beta_0g^2_{ab}(\mu)\log\frac{p^2}{\mu^2}
-2\beta_1 g_{ab}^4(\mu)\logp +\beta_0^2 g_{ab}^4(\mu)\log^2\frac{p^2}{\mu^2}\biggr]
\end{align}
We notice that the expression in square brackets is the two-loop $Z$ factor determined by the anomalous dimension of $tr F^2$ according to Eq.(\ref{eqn:dim_anomala_scalar}).
The coefficient $2\beta_1$ should become $\beta_1$ if we multiply the correlator in Eq.(\ref{eqn:corr_scalare_intermedio}) by the factor of $\bigl(1+\frac{\beta_1}{\beta_0}g^2_{ab}(\mu)\bigr)^2$, in order to make the correlator $RGI$:
\begin{align}
&\braket{\frac{\beta(g_{ab})}{Ng_{ab}}tr F^2(p)\frac{\beta(g_{ab})}{Ng_{ab}}tr F^2(-p)}_{conn}\nonumber\\
&=-\bigl(1-\frac{1}{N^2}\bigr)\beta_0^2 g^4_{ab}(\mu)
\frac{p^4}{4\pi^2}\bigl(1+\frac{\beta_1}{\beta_0}g^2_{ab}(\mu)\bigr)^2\log\frac{p^2}{\mu^2}\quad\nonumber\\
\times\quad&\biggl[1-\beta_0g^2_{ab}(\mu)\log\frac{p^2}{\mu^2}
-2\beta_1 g_{ab}^4(\mu)\logp +\beta_0^2 g_{ab}^4(\mu)\log^2\frac{p^2}{\mu^2}\biggr]\nonumber\\
&=-\bigl(1-\frac{1}{N^2}\bigr)\beta_0^2 g^4_{ab}(\mu)
\frac{p^4}{4\pi^2}\frac{\bigl(1+\frac{\beta_1}{\beta_0}g^2_{ab}(\mu)\bigr)}{\bigl(1+\frac{\beta_1}{\beta_0}g^2_{ab}(p)\bigr)}\quad\nonumber\\
\times\quad &\bigl(1+\frac{\beta_1}{\beta_0}g^2_{ab}(\mu)\bigr)\bigl(1+\frac{\beta_1}{\beta_0}g^2_{ab}(p)\bigr)\log\frac{p^2}{\mu^2}\quad\nonumber\\
\times\quad&\biggl[1-\beta_0g^2_{ab}(\mu)\log\frac{p^2}{\mu^2}
-2\beta_1 g_{ab}^4(\mu)\logp +\beta_0^2 g_{ab}^4(\mu)\log^2\frac{p^2}{\mu^2}\biggr]
\end{align}
where we have multiplied and divided by $\bigl(1+\frac{\beta_1}{\beta_0}g^2_{ab}(\mu)\bigr)$ in order to exploit the two-loop relation:
\begin{equation}
\beta_0(1+\frac{\beta_1}{\beta_0}g_{ab}^2(p))\logp =\frac{1}{g_{ab}^2(p)}-\frac{1}{g_{ab}^2(\mu)}
\end{equation}
We now evaluate separately:
\begin{align}
&\frac{(1+\frac{\beta_1}{\beta_0}g_{ab}^2(\mu))}{(1+\frac{\beta_1}{\beta_0}g_{ab}^2(p))}\nonumber\\
&=\bigl(1+\frac{\beta_1}{\beta_0}g_{ab}^2(\mu)\bigr)
\bigl(1-\frac{\beta_1}{\beta_0}g_{ab}^2(\mu)+\beta_1 g_{ab}^4(\mu)\logp +\frac{\beta_1^2}{\beta_0^2}g_{ab}^4(\mu)\bigr)+O(g^6\logp)\nonumber\\
&= 1+\beta_1 g_{ab}^4(\mu)\logp +O(g^6\logp)
\end{align}
Putting all together  we get:
\begin{align}
&\braket{\frac{\beta(g_{ab})}{Ng_{ab}}tr F^2(p)\frac{\beta(g_{ab})}{Ng_{ab}}tr F^2(-p)}_{conn}\nonumber\\
&=-\bigl(1-\frac{1}{N^2}\bigr)\beta_0 g^4_{ab}(\mu)
\frac{p^4}{4\pi^2}\quad \nonumber\\
\times\quad & \bigl(1+\beta_1 g_{ab}^4(\mu)\logp\bigr)\bigl(1+\frac{\beta_1}{\beta_0}g^2_{ab}(\mu)\bigr)
\bigl(\frac{1}{g^2_{ab}(p)}-\frac{1}{g^2_{ab}(\mu)}\bigr)\quad\nonumber\\
\times\quad&\biggl[1-\beta_0g^2_{ab}(\mu)\log\frac{p^2}{\mu^2}
-2\beta_1 g_{ab}^4(\mu)\logp +\beta_0^2 g_{ab}^4(\mu)\log^2\frac{p^2}{\mu^2}\biggr]\nonumber\\
&=-\bigl(1-\frac{1}{N^2}\bigr)\beta_0 g^4_{ab}(\mu)
\frac{p^4}{4\pi^2} \bigl(1+\frac{\beta_1}{\beta_0}g^2_{ab}(\mu)\bigr)
\bigl(\frac{1}{g^2_{ab}(p)}-\frac{1}{g^2_{ab}(\mu)}\bigr)\quad\nonumber\\
\times\quad&\biggl[1-\beta_0g^2_{ab}(\mu)\log\frac{p^2}{\mu^2}
-\beta_1 g_{ab}^4(\mu)\logp +\beta_0^2 g_{ab}^4(\mu)\log^2\frac{p^2}{\mu^2}\biggr]
\end{align}
The factor in square brackets in the last line is now precisely the renormalization factor for two powers of $g_{ab}$. Hence the correlator reads:
\begin{align}\label{corr_scalare_rgi_3_loop_improved}
&\braket{\frac{\beta(g_{ab})}{Ng_{ab}}tr F^2(p)\frac{\beta(g_{ab})}{Ng_{ab}}tr F^2(-p)}_{conn}\nonumber\\
&=-\bigl(1-\frac{1}{N^2}\bigr)\beta_0 
\frac{p^4}{4\pi^2} g^2_{ab}(\mu)g^2_{ab}(p)
\bigl(1+\frac{\beta_1}{\beta_0}g^2_{ab}(\mu)\bigr)
\bigl(\frac{1}{g^2_{ab}(p)}-\frac{1}{g^2_{ab}(p)}\bigr)\nonumber\\
&=-\bigl(1-\frac{1}{N^2}\bigr)\beta_0 
\frac{p^4}{4\pi^2} \bigl(1+\frac{\beta_1}{\beta_0}g^2_{ab}(\mu)\bigr)
\bigl(g_{ab}^2(\mu)-g_{ab}^2(p)\bigr)\nonumber\\
&=\bigl(1-\frac{1}{N^2}\bigr)\beta_0 
\frac{p^4}{4\pi^2}
\biggl[g_{ab}^2(p)\bigl(1+\frac{\beta_1}{\beta_0}g^2_{ab}(\mu)\bigr)-g_{ab}^2(\mu)\bigl(1+\frac{\beta_1}{\beta_0}g^2_{ab}(\mu)\bigr)\biggr]
\end{align}
The second term in the last line is a contact term, but the first term depends on $g_{ab}(\mu)$, therefore it is not $RGI$.
Hence Eq.(\ref{corr_scalare_rgi_3_loop_improved}) is not exactly $RGI$ even after subtracting the contact terms.
The scale dependence in the physical term is due to the fact that the correlator is not exact but it is computed to a finite order of perturbation theory.
We notice that the scale dependence occurs at order of $g^4$ only and in any case it does not affect the structure of the universal $UV$ behavior but only the overall coefficient in the $RG$ estimate.
Yet it is interesting to determine the precise overall coefficient of the asymptotic behavior. This is done for the correlator of $\frac{\beta(g)}{gN}tr F^2$ in $SU(3)$ $QCD$ in sect.(3.7) by assuming its $RG$-invariance, instead of checking it to a finite order of perturbation theory as we just did.

\subsection{Correlator of $\frac{g^2}{N} trF^2$ in $SU(N)$ $YM$ (three loops)}

We now present the result for the correlator of $\frac{g^2}{N} trF^2$. 
We recall that in this case we do not expect to get a $RGI$ function to all orders in perturbation theory.
We start from Eq.(\ref{eqn:corr_scalare_intermedio}) and we write it as:
\begin{align}\label{eqn:rg_not_improved_scalar_3l}
&\braket{\frac{g^2}{N}tr F^2(p)\frac{g^2}{N}tr F^2(-p)}_{conn}\nonumber\\
&=-\bigl(1-\frac{1}{N^2}\bigr)\frac{p^4}{4\pi^2}g^4_{ab}(\mu)
\frac{1}{\beta_0} \frac{\bigl(1+\frac{\beta_1}{\beta_0}g^2_{ab}(\mu)\bigr)}{\bigl(1+\frac{\beta_1}{\beta_0}g^2_{ab}(p)\bigr)}
\frac{1}{\bigl(1+\frac{\beta_1}{\beta_0}g^2_{ab}(\mu)\bigr)}
\biggl(\frac{1}{g^2_{ab}(p)}-\frac{1}{g^2_{ab}(\mu)}\biggr)\nonumber\\
\times\quad&\biggl[1-\beta_0g^2_{ab}(\mu)\log\frac{p^2}{\mu^2}
-2\beta_1 g_{ab}^4(\mu)\logp +\beta_0^2 g_{ab}^4(\mu)\log^2\frac{p^2}{\mu^2}\biggr]\nonumber\\
&=-\bigl(1-\frac{1}{N^2}\bigr)\frac{p^4}{4\pi^2}g^4_{ab}(\mu) 
\frac{1}{\beta_0}
\frac{1}{\bigl(1+\frac{\beta_1}{\beta_0}g^2_{ab}(\mu)\bigr)}
\biggl(\frac{1}{g^2_{ab}(p)}-\frac{1}{g^2_{ab}(\mu)}\biggr)\nonumber\\
\times\quad&\biggl[1-\beta_0g^2_{ab}(\mu)\log\frac{p^2}{\mu^2}
-\beta_1 g_{ab}^4(\mu)\logp +\beta_0^2 g_{ab}^4(\mu)\log^2\frac{p^2}{\mu^2}\biggr]\nonumber\\
&=-\bigl(1-\frac{1}{N^2}\bigr)\frac{p^4}{4\pi^2}\frac{1}{\beta_0} 
\bigl(1-\frac{\beta_1}{\beta_0}g^2_{ab}(\mu)+\frac{\beta_1^2}{\beta_0^2}g^4_{ab}(\mu)\bigr)
\bigl(g^2_{ab}(\mu)-g^2_{ab}(p)\bigr)\nonumber\\
&=\bigl(1-\frac{1}{N^2}\bigr)\frac{p^4}{4\pi^2}\frac{1}{\beta_0}\bigl(g^2_{ab}(p)-g^2_{ab}(\mu)+\frac{\beta_1}{\beta_0}g^4_{ab}(\mu)-\frac{\beta_1}{\beta_0}g^2_{ab}(p)g^2_{ab}(\mu)\bigr)
\end{align}
Surprisingly we notice that the term that depends on the product $g_{ab}(\mu)g_{ab}(p)$, that is not $RGI$, is of the same order of $g^4$ as the non-$RGI$ terms in the correlator in Eq.(\ref{corr_scalare_rgi_3_loop_improved}), that must be $RGI$.

\subsection{Correlator of $\frac{g^2}{N} trF\tilde{F}$ in $SU(N)$ $YM$ (three loops)}

We repeat the same steps to find the $RG$-improved expression for the correlator of $\frac{g^2}{N}tr F\tilde{F}$, that is $RGI$.
The three-loop correlator reads:
\begin{align}
\braket{\frac{g^2}{N}tr F\tilde{F}(p)\frac{g^2}{N}tr F\tilde{F}(-p)}_{conn}=
&-\bigl(1-\frac{1}{N^2}\bigr)\frac{p^4}{4\pi^2} \gms^4(\mu)\log\frac{p^2}{\mu^2}\quad\nonumber\\
\times\quad &\biggl[1+\gms^2(\mu)\biggl(\tilde{f}_0-\beta_0\log\frac{p^2}{\mu^2}\biggr)\nonumber\\
&+\gms^4(\mu)\biggl(\tilde{f}_1+\tilde{f}_2\log\frac{p^2}{\mu^2}+\beta_0^2\log^2\frac{p^2}{\mu^2}\biggr)\biggr]
\end{align}
Now we perform a generic change of scheme as in Eq.(\ref{eqn:cambio_schema2}):
\begin{equation}
g_{\tilde{ab}}^2=\gms^2(\mu)\bigl(1+\tilde{a}\gms^2(\mu)+\tilde{b}\gms^4(\mu)\bigr)
\end{equation}
The correlator becomes:
\begin{align}
&\braket{\frac{g^2}{N}tr F\tilde{F}(p)\frac{g^2}{N}tr F\tilde{F}(-p)}_{conn}\nonumber\\
&=-\bigl(1-\frac{1}{N^2}\bigr)\frac{p^4}{4\pi^2}g^4_{\tilde{ab}}(\mu)
\log\frac{p^2}{\mu^2}\quad\nonumber\\
\times\quad&\biggl[1+(\tilde{f}_0-2\tilde{a}) g^2_{\tilde{ab}}(\mu)
-\beta_0g^2_{\tilde{ab}}(\mu)\log\frac{p^2}{\mu^2}+(\tilde{f}_1+5\tilde{a}^2+2\tilde{b}-\tilde{a}\tilde{f}_0) g_{\tilde{ab}}^4(\mu)\nonumber\\
&+(\tilde{f}_2+3\beta_0 \tilde{a}) g_{\tilde{ab}}^4(\mu)\logp +\beta_0^2 g_{\tilde{ab}}^4(\mu)\log^2\frac{p^2}{\mu^2}\biggr]
\end{align} 
Again we impose the conditions to eliminate the finite terms:
\begin{align}
&\tilde{a}=\frac{\tilde{f}_0}{2}\nonumber\\
&\tilde{f}_1+5(\frac{\tilde{f}_0}{2})^2+2\tilde{b}-\frac{\tilde{f}_0^2}{2}=0 \nonumber\\
\Rightarrow & \tilde{b}=\frac{3}{8}\tilde{f}_0^2-\frac{\tilde{f}_1}{2}
\end{align}  
With this choice of $\tilde{a}$ the coefficient of the $g^4\logp$ term becomes:
\begin{equation}
\tilde{f}_2+3\beta_0 \tilde{a} = \tilde{f}_2+\frac{3}{2}\tilde{f}_0\beta_0=-\frac{34}{3(4\pi)^4}=-\beta_1
\end{equation}
Substituting in the correlator we get:
\begin{align}
&\braket{\frac{g^2}{N}tr F\tilde{F}(p)\frac{g^2}{N}tr F\tilde{F}(-p)}_{conn}\nonumber\\
&=-\bigl(1-\frac{1}{N^2}\bigr)\frac{p^4}{4\pi^2}g^4_{\tilde{ab}}(\mu)
\log\frac{p^2}{\mu^2}\quad \nonumber\\
\times \quad&\biggl[1-\beta_0g^2_{\tilde{ab}}(\mu)\log\frac{p^2}{\mu^2}-\beta_1 g_{\tilde{ab}}^4(\mu)\logp +\beta_0^2 g_{\tilde{ab}}^4(\mu)\log^2\frac{p^2}{\mu^2}\biggr]
\end{align}
We notice that the expression in square brackets is the two-loop $Z$ factor implied by the anomalous dimension of $trF\tilde{F}$ computed in Eq.(\ref{eqn:dim_anomala_pseudoscalar}). It renormalizes two powers of $g(\mu)$.
Therefore, the correlator reads:
\begin{align}\label{eqn:rg_improved_pseudo_3l}
&\braket{\frac{g^2}{N}tr F\tilde{F}(p)\frac{g^2}{N}tr F\tilde{F}(-p)}_{conn}\nonumber\\
&=-\bigl(1-\frac{1}{N^2}\bigr)\frac{p^4}{4\pi^2}g^2_{\tilde{ab}}(\mu) g^2_{\tilde{ab}}(p)
\log\frac{p^2}{\mu^2}\nonumber\\
&=-\bigl(1-\frac{1}{N^2}\bigr)\frac{p^4}{4\pi^2}g^2_{\tilde{ab}}(\mu) g^2_{\tilde{ab}}(p)\frac{1}{\beta_0}\biggl(\frac{1}{g_{\tilde{ab}}^2(p)}-\frac{1}{g^2_{\tilde{ab}}(\mu)}\biggr)
\frac{1}{1+\frac{\beta_1}{\beta_0} g_{\tilde{ab}}^2(p)}\nonumber\\
&=-\bigl(1-\frac{1}{N^2}\bigr)\frac{p^4}{4\pi^2}\frac{1}{\beta_0}\bigl(g^2_{\tilde{ab}}(\mu)-g^2_{\tilde{ab}}(p)\bigr)\frac{1}{1+\frac{\beta_1}{\beta_0} g_{\tilde{ab}}^2(p)}\nonumber\\
&=-\bigl(1-\frac{1}{N^2}\bigr)\frac{p^4}{4\pi^2}\frac{1}{\beta_0}\bigl(g^2_{\tilde{ab}}(\mu)-g^2_{\tilde{ab}}(p)\bigr)\bigl(1-\frac{\beta_1}{\beta_0} g_{\tilde{ab}}^2(p)+\frac{\beta_1^2}{\beta_0^2}g^4_{\tilde{ab}}(p)\bigr)\nonumber\\
&=\bigl(1-\frac{1}{N^2}\bigr)\frac{p^4}{4\pi^2}\frac{1}{\beta_0}\bigl(g^2_{\tilde{ab}}(p)+\frac{\beta_1}{\beta_0} g_{\tilde{ab}}^2(p)g^2_{\tilde{ab}}(\mu)-\frac{\beta_1}{\beta_0} g_{\tilde{ab}}^4(p)-g^2_{\tilde{ab}}(\mu)\bigr) 
\end{align}
The second term in the last line is scale dependent to the order of $g^4$ as the term that occurs in the correlator of $\frac{g^2}{N}tr F^2$ in Eq.(\ref{eqn:rg_not_improved_scalar_3l}).

\subsection{Correlator of $\frac{g^2}{N} tr {F^-}^2$ in $SU(N)$ $YM$ (three loops)}

We now sum the two results for the correlators of $trF^2$ and of $tr F\tilde{F}$ to obtain the correlator of $tr {F^-}^2$.
Indeed, we recall that:
\begin{equation}
\frac{1}{2}\braket{tr{F^-}^2(p)tr{F^-}^2(-p)}_{conn}=
2\braket{trF^2(p)trF^2(-p)}_{conn}+2\braket{trF\tilde{F}(p)trF\tilde{F}(-p)}_{conn}
\end{equation}
Summing the two results in Eq.(\ref{eqn:rg_improved_pseudo_3l}) and in Eq.(\ref{eqn:rg_not_improved_scalar_3l}) we obtain:
\begin{align}\label{eqn:corr_asd_pert}
&\frac{1}{2}\braket{\frac{g^2}{N}tr{F^-}^2(p)\frac{g^2}{N}tr{F^-}^2(-p)}_{conn}=\nonumber\\
&=\bigl(1-\frac{1}{N^2}\bigr)\frac{p^4}{2 \pi^2}\frac{1}{\beta_0} \bigl(g^2_{ab}(p) + g^2_{\tilde{ab}}(p) -g^2_{ab}(\mu) -g^2_{\tilde{ab}}(\mu) +\frac{\beta_1}{\beta_0}g^4_{ab}(\mu)\nonumber\\
&-\frac{\beta_1}{\beta_0} g_{\tilde{ab}}^4(p)+\frac{\beta_1}{\beta_0}g^2_{ab}(p)g^2_{ab}(\mu)-\frac{\beta_1}{\beta_0}g^2_{\tilde{ab}}(p)g^2_{\tilde{ab}}(\mu)\bigr)\nonumber\\
&=\bigl(1-\frac{1}{N^2}\bigr)\frac{p^4}{2\pi^2}\frac{1}{\beta_0}\bigl(g^2_{ab}(p) + g^2_{\tilde{ab}}(p) -g^2_{ab}(\mu) -g^2_{\tilde{ab}}(\mu) +\frac{\beta_1}{\beta_0}g^4_{ab}(\mu)\nonumber\\
&-\frac{\beta_1}{\beta_0} g_{\tilde{ab}}^4(p)+O(g^6)\bigr)
\end{align}
where surprisingly the mixed terms  $g^2(p)g^2(\mu)$ cancel to the order of $g^6$.
There is no perturbative explanation for such cancellation, but conjecturally the cancellation occurs because of the $RG$ invariance of the non-perturbative formula Eq.(\ref{eqn:formula_prologo}) in the $TFT$ for the $L=2$ ground state \cite{boch:crit_points} \cite{boch:glueball_prop} of the large-$N$ one-loop integrable sector of Ferretti-Heise-Zarembo (see sect.(4)).
We can express the last result in terms of the coupling constant in the $\overline{MS}$ scheme:
\begin{align}\label{eqn:corr_asd_ms}
&\frac{1}{2}\braket{\frac{g^2}{N}tr{F^-}^2(p)\frac{g^2}{N}tr{F^-}^2(-p)}_{conn}\nonumber\\
&=\bigl(1-\frac{1}{N^2}\bigr)\frac{p^4}{2\pi^2}\frac{1}{\beta_0} \bigl(2\gms^2(p)-2\gms^2(\mu) +\bigl(a+\tilde{a} -\frac{\beta_1}{\beta_0}\bigr)\gms^4(p)\nonumber\\
&+\bigl(\frac{\beta_1}{\beta_0}-a-\tilde{a}\bigr) \gms^4(\mu)\bigr)+O(g^6)
\end{align}
that coincides with Eq.(\ref{eqn:prologo:ris_sommato}).

\subsection{Scalar correlators in $SU(3)$ $QCD$ with $n_l$ massless Dirac fermions}

In this section we derive the $RG$-improved expression for the correlators of $trF^2$ and of $\frac{\beta(g_{YM})}{g_{YM}}trF^2$ in $QCD$. \par
The three-loop perturbative result for the imaginary part of the correlator of $tr F^2$ in $QCD$ with $n_l$ massless Dirac femions is \cite{chetyrkin:scalar}:
\begin{align}
&\Im{\braket{trF^2(p)trF^2(-p)}_{conn}}\nonumber\\
&=\frac{2}{\pi}p^4\Biggl\{1+\frac{\alpha_s(\mu)}{\pi}\biggl[\biggl(\frac{73}{4}-\frac{11}{2}\logp\biggr)
-n_l\biggl(\frac{7}{6}-\frac{1}{3}\logp\biggr)\biggr]\nonumber\\
&+\Bigl(\frac{\alpha_s(\mu)}{\pi}\Bigr)^2\biggr[\frac{37631}{96}-\frac{363}{8}\zeta(2)
-\frac{495}{8}\zeta(3)-\frac{2817}{16}\logp +\frac{363}{16}\log^2\frac{p^2}{\mu^2}\nonumber\\
&+n_l\biggl(-\frac{7189}{144}+\frac{11}{2}\zeta(2)+\frac{5}{4}\zeta(3)+\frac{263}{12}\logp-\frac{11}{4}\log^2\frac{p^2}{\mu^2}\biggr)\nonumber\\
&+n_l^2\biggl(\frac{127}{108}-\frac{1}{6}\zeta(2)-\frac{7}{12}\logp +\frac{1}{12}\log^2\frac{p^2}{\mu^2}\biggr)\biggr]\Biggr\}
\end{align}
We write the correlator in terms of the coupling $g_{YM}$ in the $\overline{MS}$ scheme instead of $\alpha_s$:
\begin{align}\label{eqn:im_pert}
&\Im{\braket{trF^2(p)trF^2(-p)}_{conn}}\nonumber\\
&=\frac{2}{\pi}p^4\Biggl\{1+g_{YM}^2(\mu)\biggl[\biggl(73-22\logp\biggr)
-n_l\biggl(\frac{14}{3}-\frac{4}{3}\logp\biggr)\biggr]\frac{1}{(4\pi)^2}\nonumber\\
&+g_{YM}^4(\mu)\biggr[\biggl(\frac{37361}{6}-726\zeta(2)-990\zeta(3)
-2817\logp +363\log^2\frac{p^2}{\mu^2}\biggr)\nonumber\\
&+n_l\biggl(-\frac{7189}{9}+88\zeta(2)+20\zeta(3)+\frac{1052}{3}\logp-44\log^2\frac{p^2}{\mu^2}\biggr)\nonumber\\
&+n_l^2\biggl(\frac{508}{27}-\frac{8}{3}\zeta(2)-\frac{28}{3}\logp +\frac{4}{3}\log^2\frac{p^2}{\mu^2}\biggr)\biggr]\frac{1}{(4\pi)^4}\Biggr\}
\end{align}
If we suppose the correlator to be of the form:
\begin{align}
&\braket{trF^2(p)trF^2(-p)}_{conn}\nonumber\\
&=-\frac{2}{\pi^2}p^4\logp
\biggl[1+g_{YM}^2(\mu)\Bigl(h_0+h_1\logp\Bigr)\nonumber\\
&+g_{YM}^4(\mu)\Bigl(h_2+h_3\logp +h_4 \log^2\frac{p^2}{\mu^2}\Bigr)\biggr]
\end{align}
its imaginary part is:
\begin{align}\label{eqn:im_ipotesi}
&\Im{\braket{trF^2(p)trF^2(-p)}_{conn}}\nonumber\\
&=\frac{2}{\pi}p^4\biggl[1+h_0 g_{YM}^2(\mu) + 2h_1 g_{YM}^2(\mu)\logp \nonumber\\
&+(h_2-\pi^2 h_4)g_{YM}^4(\mu) +2h_3 g_{YM}^4(\mu)\logp +3h_4 g_{YM}^4(\mu)\log^2\frac{p^2}{\mu^2}\biggr]
\end{align}
Comparing Eq.(\ref{eqn:im_ipotesi}) and Eq.(\ref{eqn:im_pert}) we get:
\begin{align}
h_0&=\biggl(73-\frac{14}{3}n_l\biggr)\frac{1}{(4\pi)^2}\nonumber\\
2h_1&=\biggl(-22+\frac{4}{3}n_l\biggr)\frac{1}{(4\pi)^2}\nonumber\\
\Rightarrow h_1&=\biggl(-11+\frac{2}{3}n_l\biggr)\frac{1}{(4\pi)^2}=-\tilde{\beta}_0\nonumber\\
h_2-\pi^2 h_4 &=\biggl[\frac{37361}{6}-726\zeta(2)
-990\zeta(3)
+n_l\Bigl(-\frac{7189}{9}+88\zeta(2)+20\zeta(3)\Bigr)\nonumber\\
&+n_l^2\Bigl(\frac{508}{27}-\frac{8}{3}\zeta(2)\Bigr)\biggr]\frac{1}{(4\pi)^4}\nonumber\\
2h_3 &=\biggl[-2817+\frac{1052}{3}n_l-\frac{28}{3}n_l^2\biggr]\frac{1}{(4\pi)^4}\nonumber\\
\Rightarrow h_3 &=
\biggl[-\frac{2817}{2}+\frac{526}{3}n_l-\frac{14}{3}n_l^2\biggr]\frac{1}{(4\pi)^4}\nonumber\\
3h_4&=\biggl[363- 44n_l+\frac{4}{3}n_l^2\biggr]\frac{1}{(4\pi)^4}\nonumber\\
\Rightarrow h_4&=\biggl[121- \frac{44}{3}n_l+\frac{4}{9}n_l^2\biggr]\frac{1}{(4\pi)^4}=\tilde{\beta}_0^2
\end{align}
Now we repeat the same steps as in the $n_l=0$ case. 
We change renormalization scheme in order to cancel the finite parts:
\begin{equation}
g_{uv}^2(\mu)=g_{YM}^2(\mu)\bigl(1+u g_{YM}^2(\mu)+vg_{YM}^4(\mu)\bigr)
\end{equation} 
We use the perturbative expression for the renormalized coupling constant with two-loop accuracy:
\begin{align*}
&g_{YM}^2(p)=g_{YM}^2(\mu)\Bigl(1-\tilde{\beta}_0 g_{YM}^2(\mu)\logp -\tilde{\beta}_1 g_{YM}^4(\mu)\logp\nonumber\\ &+\tilde{\beta}_0^2g^4_{YM}(\mu)\log^2\frac{p^2}{\mu^2}\Bigr)
\end{align*} 
where the tilde refers to the $QCD$ coefficients of the $\beta$ function: 
\begin{align}
\tilde{\beta}_0&=\Bigl(11-\frac{2}{3}n_l\Bigr)\frac{1}{(4\pi)^2}\nonumber\\
\tilde{\beta}_1 &=\Bigl(102-\frac{38}{3}n_l\Bigr)\frac{1}{(4\pi)^4}
\end{align}
\par
We consider now the correlator of $g_{YM}^2 tr F^2$:
\begin{align}
&\braket{g_{YM}^2 trF^2(p)g_{YM}^2trF^2(-p)}_{conn}\nonumber\\
&=-\frac{2g_{YM}^4(\mu)}{\pi^2}p^4\logp
\biggl[1+g_{YM}^2(\mu)\Bigl(h_0-\tilde{\beta}_0\logp\Bigr)\nonumber\\
&+g_{YM}^4(\mu)\Bigl(h_2+h_3\logp +\tilde{\beta}_0^2 \log^2\frac{p^2}{\mu^2}\Bigr)\biggr]\nonumber\\
&=-\frac{2g^4_{uv}(\mu)}{\pi^2}p^4
\log\frac{p^2}{\mu^2}\quad \nonumber\\
\times\quad&\biggl[1+(h_0-2u) g^2_{uv}(\mu)-
\tilde{\beta}_0 g^2_{uv}(\mu)\log\frac{p^2}{\mu^2}+(h_2+5u^2+2v-uh_0) g_{uv}^4(\mu)\nonumber\\
&+(h_3+3\tilde{\beta}_0 u) g_{uv}^4(\mu)\logp +\tilde{\beta}_0^2 g_{uv}^4(\mu)\log^2\frac{p^2}{\mu^2}\biggr]
\end{align}
Choosing $u=\frac{h_0}{2}$ to cancel the finite term of order of $g^2$ in the square brackets we get for the coefficient of the term of order of $g^4 \logp$:
\begin{align}
&h_3+3\tilde{\beta}_0 u\nonumber\\
&=h_3+\frac{3}{2}\tilde{\beta}_0 h_0\nonumber\\
&=\biggl(-\frac{2817}{2}+\frac{526}{3}n_l-\frac{14}{3}n_l^2\biggr)\frac{1}{(4\pi)^4}
+\frac{3}{2}\biggl(73-\frac{14}{3}n_l\biggr)\biggl(11-\frac{2}{3}n_l\biggr)\frac{1}{(4\pi)^4}\nonumber\\
&=\biggl(-\frac{2817}{2}+\frac{526}{3}n_l-\frac{14}{3}n_l^2+\frac{2409}{2}-73n_l-77n_l+\frac{14}{3}n_l^2\biggr)\frac{1}{(4\pi)^4}\nonumber\\
&=-204+\frac{76}{3}n_l\nonumber\\
&=-2\tilde{\beta}_1
\end{align}
as predicted by Eq.(\ref{eqn:dim_anomala_scalar}) and by the computational experience gained in the pure $YM$ case. 
To cancel the finite term of order of $g^4$ we put:
\begin{equation}
h_2+\frac{5}{2}h_0^2+2v-\frac{h_0^2}{2}=0
\end{equation} 
Therefore, the correlator reads:
\begin{align}\label{eqn:corr_scalare_qcd}
&\braket{g_{YM}^2trF^2(p)g_{YM}^2trF^2(-p)}_{conn}\nonumber\\
&=-\frac{2g^4_{uv}(\mu)}{\pi^2}p^4
\log\frac{p^2}{\mu^2}\biggl[1-\tilde{\beta}_0 g^2_{uv}(\mu)\log\frac{p^2}{\mu^2}-2\tilde{\beta}_1 g_{uv}^4(\mu)\logp \nonumber\\
&+\tilde{\beta}_0^2 g_{uv}^4(\mu)\log^2\frac{p^2}{\mu^2}\biggr]
\end{align}
Now we follow the same steps as in the $n_l=0$ case. The only differences are the coefficients of the $\beta$ function and the parameters $u,v$ that define the new renormalization scheme.
The result is:
\begin{align}
&\braket{g_{YM}^2trF^2(p)g_{YM}^2trF^2(-p)}\nonumber\\
&=\frac{2}{\tilde{\beta}_0\pi^2}p^4\Bigl(g^2_{uv}(p)-\frac{\tilde{\beta}_1}{\tilde{\beta}_0}g^2_{uv}(p)g^2_{uv}(\mu)-g^2_{uv}(\mu)+\frac{\tilde{\beta}_1}{\tilde{\beta}_0}g^4_{uv}(\mu)\Bigr)
\end{align}
\newline
Hence:
\begin{align}
\biggl(\frac{\beta(g_{uv})}{g_{uv}}\biggr)^2 \Pi(\frac{p}{\mu})
=\frac{2\tilde{\beta}_0}{\pi^2}
\biggl(g_{uv}^2(p)+\frac{\tilde{\beta}_1}{\tilde{\beta}_0}g^2_{uv}(\mu)g_{uv}^2(p)-g_{uv}^2(\mu)
-\frac{\tilde{\beta}_1}{\tilde{\beta}_0}g^4_{uv}(\mu)\biggr)
\end{align}
We recall that $u=\frac{h_0}{2}$, therefore:
\begin{equation}
g_{uv}^2(\mu)=g_{YM}^2(\mu)\bigl(1+\frac{h_0}{2}g_{YM}^2(\mu)+vg_{YM}^4(\mu)\bigr)
\end{equation}
Hence we get:
\begin{align}\label{eqn:ris_vecchio}
&\biggl(\frac{\beta(g_{uv})}{g_{uv}}\biggr)^2 \Pi(\frac{p}{\mu})\nonumber\\
&=\frac{2\tilde{\beta}_0}{\pi^2}
\biggl(g_{YM}^2(p)+\frac{h_0}{2}g_{YM}^4(p)+\frac{\tilde{\beta}_1}{\tilde{\beta}_0}g^2_{YM}(\mu)g_{YM}^2(p)+\nonumber\\
&-g_{YM}^2(\mu)
-\frac{h_0}{2}g_{YM}^4(\mu)-\frac{\tilde{\beta}_1}{\tilde{\beta}_0}g^4_{YM}(\mu)+O(g^6)\biggr)
\end{align}
Now we  multiply the $RHS$ of Eq.(\ref{eqn:ris_vecchio}) by 
$\frac{\bigl(\frac{\beta(g_{YM})}{g_{YM}}\bigr)^2}{\bigl(\frac{\beta(g_{uv})}{g_{uv}}\bigr)^2}$.
Indeed, this is necessary to take into account the change of scheme performed to compute Eq.(\ref{eqn:ris_vecchio}).
The additional factor is:
\begin{align}
&\frac{\bigl(\frac{\beta(g_{YM})}{g_{YM}}\bigr)^2}{\bigl(\frac{\beta(g_{uv})}{g_{uv}}\bigr)^2}=
\frac{\bigl(1+\frac{\beta_1}{\beta_0}g_{YM}^2(\mu)\bigr)^2}{(1+\frac{\beta_1}{\beta_0}g_{uv}^2(\mu)\bigr)^2}\nonumber\\
&=\bigl(1+2\frac{\beta_1}{\beta_0}g_{YM}^2(\mu)+\frac{\beta_1^2}{\beta_0^2}g_{YM}^4(\mu)\bigr)
\bigl(1-2\frac{\beta_1}{\beta_0}g_{uv}^2(\mu)+3\frac{\beta_1^2}{\beta_0^2}g_{YM}^4(\mu)\bigr)\nonumber\\
&=1-h_0\frac{\beta_1}{\beta_0}g^4_{YM}(\mu) +O(g^6)
\end{align}
Therefore, the correlator in Eq.(\ref{eqn:ris_vecchio}) becomes:
\begin{align}
&\biggl(\frac{\beta(g_{YM})}{g_{YM}}\biggr)^2 \Pi(\frac{p}{\mu})\nonumber\\
&=\frac{2\tilde{\beta}_0}{\pi^2}
\biggl(g_{YM}^2(p)+\frac{h_0}{2}g_{YM}^4(p)+\frac{\tilde{\beta}_1}{\tilde{\beta}_0}g^2_{YM}(\mu)g_{YM}^2(p)+\nonumber\\
&-g_{YM}^2(\mu)
-\frac{h_0}{2}g_{YM}^4(\mu)-\frac{\tilde{\beta}_1}{\tilde{\beta}_0}g^4_{YM}(\mu)-h_0\frac{\beta_1}{\beta_0}g^4_{YM}(\mu)+O(g^6)\biggr)
\end{align}
that has some dependence on the scale $\mu$ even after subtracting the contact terms. In the next section we get rid of this dependence 
by using a different method, that assumes the $RG$ invariance of the correlator instead of checking it. \par
In any case the universal $UV$ asymptotic behavior is in agreement with the $RG$ estimate, i.e.:
\begin{align}
&\braket{\frac{\beta(g_{YM})}{g_{YM}}tr F^2(p)\frac{\beta(g_{YM})}{g_{YM}}tr F^2(-p)}_{conn}
\sim \frac{p^4}{\tilde{\beta}_0 \log\plms}\Biggl(1-\frac{\tilde{\beta}_1}{\tilde{\beta}_0^2}\frac{\log\log\plms}{\log\plms}\Biggr)
\end{align}

\subsection{$RG$-invariant scalar correlator in $SU(3)$ $QCD$ with $n_l$ massless Dirac fermions}

Firstly, we check the correctness of the finite parts of the scalar correlator in $QCD$, reconstructed in sect.(3.6) from its imaginary part, thanks to another result reported in \cite{chet:tensore}:
\begin{align}\label{eqn:der_chet}
p^2\frac{d}{d p^2} \Pi(p)\biggl|_{\logp=0}=
\frac{1}{\pi^2}\biggl[-2+\frac{\alpha_s}{\pi}\biggl(-\frac{73}{2}+\frac{7}{3}n_l\biggr)+\nonumber\\
+\frac{\alpha_s^2}{\pi^2}\biggl(-\frac{37631}{48}+\frac{495}{4}\zeta(3)+n_l\Bigl(\frac{7189}{72}-\frac{5}{2}\zeta(3)\Bigr)-\frac{127}{54}n_l^2\biggr)\biggr]
\end{align}
with:
\begin{equation}
p^4 \Pi(p)=\braket{trF^2(p)trF^2(-p)}
\end{equation}
where we have changed the overall normalization factor of the correlator with respect to \cite{chet:tensore} to be coherent with the one used in this paper.
We perform the derivative at $p^2=\mu^2$ of the correlator obtained in sect.(3.6):
\begin{align}\label{eqn:correlator}
&p^2\frac{d}{d p^2} \Pi(\frac{p}{\mu})\biggl|_{\logp=0}\nonumber\\
&=-\frac{d}{d\log p^2}\biggl|_{\logp=0}\biggl[ \frac{2}{\pi^2}\logp
\biggl(1+g_{YM}^2(\mu)\Bigl(h_0+h_1\logp\Bigr)\nonumber\\
&+g_{YM}^4(\mu)\Bigl(h_2+h_3\logp +h_4 \log^2\frac{p^2}{\mu^2}\Bigr)\biggr)\biggr] \nonumber\\
&=-\frac{2}{\pi^2}\biggl(1+g_{YM}^2(\mu) h_0+g_{YM}^4(\mu) h_2\biggr) 
\end{align}
We recall that:
\begin{align}\label{eqn:der_mia}
h_0&=\biggl(73-\frac{14}{3}n_l\biggr)\frac{1}{(4\pi)^2}\nonumber\\
h_2&=\frac{37361}{6}-990\zeta(3) + n_l\Bigl(-\frac{7189}{9}+20\zeta(3)\Bigr)+\frac{508}{27}n_l^2
\end{align}
It is easy to verify that Eq.(\ref{eqn:der_chet}) and Eq.(\ref{eqn:der_mia}) are in agreement.
Indeed:
\begin{align}
&\frac{1}{\pi^2}\biggl[-2+\frac{\alpha_s}{\pi}\biggl(-\frac{73}{2}+\frac{7}{3}n_l\biggr)\nonumber\\
&+\frac{\alpha_s^2}{\pi^2}\biggl(-\frac{37631}{48}+\frac{495}{4}\zeta(3)+n_l\Bigl(\frac{7189}{72}-\frac{5}{2}\zeta(3)\Bigr)-\frac{127}{54}n_l^2\biggr)\biggr]\nonumber\\
&=-\frac{2}{\pi^2}\biggl[1+\frac{g_{YM}^2}{4\pi^2}\biggl(+\frac{73}{4}-\frac{7}{6}n_l\biggr)\nonumber\\
&+\frac{g_{YM}^4}{(4\pi^2)^2}\biggl(+\frac{37631}{96}-\frac{495}{8}\zeta(3)+n_l\Bigl(-\frac{7189}{144}+\frac{5}{4}\zeta(3)\Bigr)+\frac{127}{108}n_l^2\biggr)\biggr]\nonumber\\
&=-\frac{2}{\pi^2}\biggl[1+g_{YM}^2h_0+g_{YM}^4 h_2\biggr]
\end{align}
From Eq.(\ref{eqn:correlator}) it follows the derivative of the correlator of $\frac{\beta(g_{YM})}{g_{YM}} trF^2$ with two-loop accuracy:
\begin{align}\label{eqn:der_corr_rgi}
&p^2\frac{d}{d p^2} \biggl(\frac{\beta(g_{YM})}{g_{YM}}\biggr)^2\Pi(\frac{p}{\mu})\biggl|_{\logp=0}\nonumber\\
&=-\frac{2}{\pi^2}\tilde{\beta}_0^2 g^4_{YM}(\mu)\bigl(1+\frac{\tilde{\beta}_1}{\tilde{\beta}_0}g_{YM}^2(\mu)\bigr)^2\biggl[1+g_{YM}^2(\mu)h_0+g_{YM}^4(\mu) h_2\biggr]
\end{align}
Secondly, we write $g_{YM}(p)$ instead of $g_{YM}(\mu)$  in Eq.(\ref{eqn:der_corr_rgi}) since $\logp=0\Rightarrow p^2=\mu^2$ in order to get 
the large-momentum correlator in a manifestly $RGI$ form.
Exploiting the definition of the $\beta$ function we can express $d\log p^2$ in terms of $dg(p)$:
\begin{align}
&\frac{dg}{d\log p}=\beta(g)
\Rightarrow  d\log(p^2)=2\frac{dg}{\beta(g)} = \frac{d(g^2)}{g\beta(g)}
\end{align} 
We integrate Eq.(\ref{eqn:der_corr_rgi}) to obtain:
\begin{align}\label{eqn:ris_nuovo}
&\frac{d}{d \log p^2} \biggl(\frac{\beta(g)}{g}\biggr)^2\Pi(\frac{p}{\mu})\biggl|_{\logp=0}\nonumber\\
&=-\frac{2}{\pi^2}\tilde{\beta}_0^2 g^4_{YM}(p)\bigl(1+\frac{\tilde{\beta}_1}{\tilde{\beta}_0}g_{YM}^2(p)\bigr)^2\biggl[1+g_{YM}^2(p)h_0+g_{YM}^4(p) h_2\biggr] \nonumber\\
\Rightarrow & \biggl(\frac{\beta(g_{YM})}{g_{YM}}(p)\biggr)^2 \Pi(\frac{p}{\mu}) - \biggl(\frac{\beta(g_{YM})}{g_{YM}}(\mu)\biggr)^2\Pi(1) \nonumber\\
&=\frac{2}{\pi^2}\tilde{\beta}_0^2 \int_{g_{YM}^2(\mu)}^{g^2_{YM}(p)}g_{YM}^4\bigl(1+\frac{\tilde{\beta}_1}{\tilde{\beta}_0}g_{YM}^2\bigr)^2\biggl[1+g_{YM}^2h_0+g_{YM}^4 h_2\biggr]\frac{d(g_{YM}^2)}{\tilde{\beta_0}g_{YM}^4(1+\frac{\tilde{\beta_1}}{\tilde{\beta_0}}g_{YM}^2)}\nonumber\\
&=\frac{2}{\pi^2}\tilde{\beta}_0\biggl[g_{YM}^2(p)-g_{YM}^2(\mu)+\biggl(\frac{\tilde{\beta}_1}{2\tilde{\beta}_0}+\frac{h_0}{2}\biggr)\biggl(g_{YM}^4(p)-g_{YM}^4(\mu)\biggr)+ O(g^6)\biggr]
\end{align}
Eq.(\ref{eqn:ris_nuovo}) gives the manifestly $RGI$ form of the correlator after subtracting the $\mu$-dependent contact terms.

\subsection{Correlators in the coordinate representation}

In this section we find the $RG$-improved expression for the perturbative correlators in the coordinate representation.
This procedure has the main advantage that in the coordinate representation the contact terms do not occur, since they are eliminated by the Fourier transform.
Indeed, the Fourier transform of $p^4$ is:
\begin{equation}
\int p^4  e^{ip\cdot x} \frac{d^4 p}{(2\pi)^4}=
\Delta^2 \delta (x) 
\end{equation}
that is supported only at $x=0$. This implies that at points different from zero the contact terms do not occur. 
The $RG$ improvement and the Fourier transform must commute up to perhaps finite scheme-dependent terms. Therefore, in this way we get another check of the asymptotic behavior.
In appendix A we compute the Fourier transforms necessary to pass from the momentum to the coordinate representation.
In particular we use the following results:
\begin{align}
\int {(p^2)}^2\log\frac{p^2}{\mu^2} e^{ip \cdot x}  \frac{d^4p}{(2\pi)^4} &=
-\frac{2^6\cdot 3}{\pi^2 x^8}\nonumber\\
\int {(p^2)}^2\biggl(\log\frac{p^2}{\mu^2}\biggr)^2 e^{ip \cdot x}  \frac{d^4p}{(2\pi)^4} &=
\frac{2^7\cdot 3}{\pi^2 x^8}\bigl(-\frac{10}{3}+2\gamma_E -\log\frac{4}{x^2\mu^2}\bigr)\nonumber\\
\int {(p^2)}^2\biggl(\log\frac{p^2}{\mu^2}\biggr)^3 e^{ip \cdot x}  \frac{d^4p}{(2\pi)^4} &=
\frac{2^6\cdot 3}{\pi^2 x^8}\bigl(-\frac{51}{2}+40\gamma_E-12\gamma_E^2\nonumber\\
&-(20-12\gamma_E)\log\frac{4}{x^2\mu^2} -3\log^2\frac{4}{x^2\mu^2}\bigr)\nonumber\\
\end{align}
Using these formulae to compute the Fourier transform of the two-loop perturbative result in Eq.(\ref{eqn:2l_perturbative_corr}) we get, disregarding the finite parts in Eq.(\ref{eqn:2l_perturbative_corr}):
\begin{align}\label{eqn:int_corr_2l}
&-\int \gms^4(\mu)p^4\logp\biggl[1-\beta_0 \gms^2(\mu)\logp\biggr]e^{ip\cdot x}\frac{d^4p}{(2\pi)^4}\nonumber\\
&=\frac{3\cdot 2^6}{\pi^2 x^8} \gms^4(\mu)
+\beta_0 \gms^6(\mu)\frac{3\cdot 2^6}{\pi^2 x^8}\bigl[-2\log\frac{4}{x^2\mu^2} +4\gamma_E - \frac{20}{3}\bigr]\nonumber\\
&= \frac{3\cdot 2^6}{\pi^2 x^8}\gms^4(\mu)\biggl[1+\bigl(-\beta_0\frac{20}{3}+4\beta_0\gamma_E\bigr) \gms^2(\mu)-2\beta_0 \gms^2(\mu)\log\frac{4}{x^2\mu^2}\biggr]
\end{align}
Firstly, the Fourier transform has produced a new finite part. 
Secondly, the coefficient of the logarithm in the square brackets is multiplied by two after the Fourier transform. This implies that the factor in the square brackets renormalizes four powers of $\gms(\mu)$, as opposed to the momentum representation, where only two powers of the coupling constant were renormalized. 
This is as expected, since in the coordinate representation the correlator is multiplicatively renormalizable as implied by Eq.(\ref{eqn:pert_general_behavior_x}). \par
To eliminate the finite term arising from the Fourier transform we change scheme defining:
\begin{equation}
g^2_s(\mu)= g^2(\mu)\bigl[1+\frac{1}{2}\bigl(-\beta_0\frac{20}{3}+4\beta_0\gamma_E\bigr)g^2(\mu)\bigr]
\end{equation} 
The integral in Eq.(\ref{eqn:int_corr_2l}) reads:
\begin{align}
&-\int \gms^4(\mu)p^4\logp\biggl[1-\beta_0 \gms^2(\mu)\logp\biggr]e^{ip\cdot x}\frac{d^4p}{(2\pi)^4}=
\frac{3\cdot 2^6}{\pi^2 x^8} g^4(x)\nonumber\\
\end{align} 
where $g(x)$ is the one-loop running coupling in the coordinate scheme \cite{chetyrkin:TF}:
\begin{align}
g^2(x)=g^2(\mu)\biggl[1-\beta_0 g^2(\mu)\log\frac{4}{x^2\mu^2}\biggr]
\end{align}
Therefore, the renormalization group improved one-loop asymptotic expression for the correlator is:
\begin{equation}\label{eqn:rg_improved_2l_{a}}
\braket{\frac{g^2}{N} trF^2(x) \frac{g^2}{N} tr F^2(0)}_{conn}\sim
 \bigl(1-\frac{1}{N^2}\bigr) \frac{3\cdot 2^6}{\pi^2 x^8} \frac{1}{\log^2\frac{4}{x^2\mu^2}}
\end{equation}
The Fourier transform provides automatically the change in sign necessary to obtain a positive expression.
This is due to the fact that in the coordinate representation contact terms do not occur.
We now go one step further performing the Fourier transform of the three-loop propagators in Eq.(\ref{eqn:corr_pert_scalar_3l}) and in Eq.(\ref{eqn:corr_pert_pseudoscalar_3l}). 
We start with the scalar correlator up to the overall normalization:
 \begin{align}\label{eqn:int_corr_3l}
&-\int g_{ab}^4(\mu)p^4\logp \nonumber\\
&\times\biggl[1-\beta_0 (\mu)g^2_{ab}(\mu)\logp -2\beta_1 g^4_{ab}(\mu)\logp +\beta_0^2 g^4_{ab}(\mu)\log^2\frac{p^2}{\mu^2}\biggr]e^{ip\cdot x}\frac{d^4p}{(2\pi)^4}\nonumber\\
&=\frac{2^6\cdot 3}{\pi^2 x^8}g_{ab}^4(\mu)\biggl[1+\bigl(-\beta_0\frac{20}{3}+4\beta_0\gamma_E\bigr)g^2_{ab}(\mu)
-2\beta_0 g^2_{ab}(\mu)\log\frac{4}{x^2\mu^2} \nonumber\\
&+\bigl(8\beta_1\gamma_E-\frac{40}{3}\beta_1+\frac{51}{2}\beta_0^2 -40\beta_0^2\gamma_E+12\beta_0^2\gamma_E^2\bigr)g_{ab}^4(\mu)
-4\beta_1 g_{ab}^4(\mu)\log\frac{4}{x^2\mu^2} \nonumber\\
&-\beta_0^2\bigl(12\gamma_E-20\bigr)g^4_{ab}(\mu)\log\frac{4}{x^2\mu^2}
+3\beta_0^2 g^4_{ab}(\mu)\log^2\frac{4}{x^2\mu^2}\biggr]
\end{align}
The following scheme redefinition:
\begin{equation}
g^2_{st}(\mu)= g^2_{ab}(\mu)\bigl(1+(2\beta_0\gamma_E-\frac{10}{3}\beta_0)g^2_{ab}+t g_{ab}^4(\mu)\bigr)
\end{equation}
cancels the finite term of order of $g^2$ in the square brackets and some terms of order of $g^4\log\frac{4}{x^2\mu^2}$, leaving only the term proportional to $-4\beta_1$.
Moreover, the finite term of order of $g^4$ in the square brackets is cancelled by a suitable choice of $t$, as in the previous section.
Eq.(\ref{eqn:int_corr_3l}) now reads:
\begin{align}\label{eqn:corr_scalare_3l_revisited}
&-\int g_{ab}^4(\mu)p^4\logp \nonumber\\
&\times\biggl[1-\beta_0 (\mu)g^2_{ab}(\mu)\logp -2\beta_1 g^4_{ab}(\mu)\logp +\beta_0^2 g^4_{ab}(\mu)\log^2\frac{p^2}{\mu^2}\biggr]e^{ip\cdot x}\frac{d^4p}{(2\pi)^4}\nonumber\\
&=\frac{2^6\cdot 3}{\pi^2 x^8}g_{st}^4(\mu)
\frac{1+2\frac{\beta_1}{\beta_0}g^2_{st}(\mu)}{1+2\frac{\beta_1}{\beta_0}g^2_{st}(x)}
\frac{1+2\frac{\beta_1}{\beta_0}g^2_{st}(x)}{1+2\frac{\beta_1}{\beta_0}g^2_{st}(\mu)}\quad \nonumber\\
\times\quad &\biggl[1-2\beta_0 g^2_{st}(\mu)\log\frac{4}{x^2\mu^2}
-4\beta_1 g_{st}^4(\mu)\log\frac{4}{x^2\mu^2}+3\beta_0^2 g^4_{st}(\mu)\log^2\frac{4}{x^2\mu^2}\biggr]\nonumber\\
&=\frac{2^6\cdot 3}{\pi^2 x^8}g_{st}^4(\mu)
\frac{1+2\frac{\beta_1}{\beta_0}g^2_{st}(x)}{1+2\frac{\beta_1}{\beta_0}g^2_{st}(\mu)}\quad \nonumber\\
\times\quad & \biggl[1-2\beta_0 g^2_{st}(\mu)\log\frac{4}{x^2\mu^2}
-2\beta_1 g_{st}^4(\mu)\log\frac{4}{x^2\mu^2}+3\beta_0^2 g^4_{st}(\mu)\log^2\frac{4}{x^2\mu^2}\biggr]\nonumber\\
&=\frac{2^6\cdot 3}{\pi^2 x^8}g_{st}^4(x)\Bigl(1+2\frac{\beta_1}{\beta_0}g^2_{st}(x)
-2\frac{\beta_1}{\beta_0}g^2_{st}(\mu)\Bigr)
\end{align}
The scale dependent term in Eq.(\ref{eqn:corr_scalare_3l_revisited}) occurs now at the order of $g^6$, while in the  momentum representation occurred at the order of $g^4$.
Now we multiply both sides of Eq.(\ref{eqn:corr_scalare_3l_revisited}) by $\bigl(1+\frac{\beta_1}{\beta_0}g_{st}^2(\mu)\bigr)^2$, i.e. by the factor necessary to make the correlator $RGI$.
Reinserting the overall normalization, we obtain:
\begin{align}
&\int \braket{\frac{\beta(g)}{Ng} trF^2(p)\frac{\beta(g)}{Ng}trF^2(-p)}_{conn}e^{ip\cdot x}\frac{d^4p}{(2\pi)^4}\nonumber\\ 
&= \frac{1}{4 \pi^2} \bigl(1-\frac{1}{N^2}\bigr)\frac{2^6\cdot 3}{\pi^2 x^8}g_{st}^4(x)\Bigl(1+\frac{\beta_1}{\beta_0}g_{st}^2(\mu)\Bigr)^2\Bigl(1+2\frac{\beta_1}{\beta_0}g^2_{st}(x)
-2\frac{\beta_1}{\beta_0}g^2_{st}(\mu)\Bigr)\nonumber\\
&= \frac{1}{4 \pi^2}\bigl(1-\frac{1}{N^2}\bigr)\frac{2^6\cdot 3}{\pi^2 x^8}g_{st}^4(x)\Bigl(1+2\frac{\beta_1}{\beta_0}g^2_{st}(x) +O(g^4)\Bigr)
\end{align}
As a result the possible scale dependence is of order of $g^8$. \par
Performing the same steps for the pseudoscalar correlator in Eq.(\ref{eqn:corr_pert_pseudoscalar_3l}) we get:
\begin{align}\label{eqn:corr_pseudoscalare_3l_revisited}
&\int \braket{\frac{g^2}{N}trF\tilde{F}(p) \frac{g^2}{N}trF\tilde{F}(-p)}_{conn}e^{ip\cdot x}\frac{d^4p}{(2\pi)^4}\nonumber\\
&=\frac{1}{4 \pi^2}\bigl(1-\frac{1}{N^2}\bigr) \frac{2^6\cdot 3}{\pi^2 x^8}g_{\tilde{st}}^4(\mu)\biggl[1-2\beta_0 g^2_{\tilde{st}}(\mu)\log\frac{4}{x^2\mu^2}
-2\beta_1 g^4_{\tilde{st}}(\mu)\log\frac{4}{x^2\mu^2}+3\beta_1^2g^4_{\tilde{st}}\log^2\frac{4}{x^2\mu^2}\biggr] \nonumber\\
&= \frac{1}{4 \pi^2}\bigl(1-\frac{1}{N^2}\bigr)\frac{2^6\cdot 3}{\pi^2 x^8}g_{\tilde{st}}^4(x) 
\end{align}
The correlator in Eq.(\ref{eqn:corr_pseudoscalare_3l_revisited}) is $RGI$ in the coordinate representation with three-loop accuracy, while in the momentum representation scale-dependent terms of the order of $g^4$ occurred in Eq.(\ref{eqn:rg_improved_pseudo_3l}). 
As in the momentum representation we find the correlator of $tr{F^-}^2$ summing the double of the scalar Eq.(\ref{eqn:corr_scalare_3l_revisited}) and pseudoscalar Eq.(\ref{eqn:corr_pseudoscalare_3l_revisited}) correlators.
We obtain:
\begin{align}\label{eqn:corr_asd_{a}}
&\frac{1}{2}\int \braket{\frac{g^2}{N}tr{F^-}^2(p)\frac{g^2}{N}tr{F^-}^2(-p)}_{conn}e^{ip\cdot x}\frac{d^4p}{(2\pi)^4}\nonumber\\
&=\frac{1}{4 \pi^2} \bigl(1-\frac{1}{N^2}\bigr)\frac{2^7\cdot 3}{\pi^2 x^8}\bigl(g^4_{st}(x)+g^4_{\tilde{st}}(x)+2\frac{\beta_1}{\beta_0}g^6_{st}(x)-2\frac{\beta_1}{\beta_0}g^2_{st}(\mu)g^4_{st}(x)\bigr)\nonumber\\
\end{align}
The scale dependence enters the term of order of $g^6$ as in the momentum representation in Eq.(\ref{eqn:corr_asd_pert}). \par
We check the correctness of the separation of the contact terms performed in the momentum representation.
We verify to the order of the leading logarithm that the Fourier transform of the $RG$-improved expression in the momentum representation in Eq.(\ref{eqn:corr_asd_ms}) is equal to Eq.(\ref{eqn:corr_asd_{a}}) in the coordinate representation.
Within the leading logarithmic accuracy it is sufficient to put $g^2(p)$:
\begin{equation}
g^2(p)=\frac{g^2(\mu)}{1+\beta_0 g^2(\mu)\logp}
\end{equation}
Therefore, the Fourier transform of the correlator in Eq.(\ref{eqn:corr_asd_ms}) can be computed reducing it to a series of positive powers of logarithms:
\begin{align}\label{eqn:serie_p_da_trasformare}
& \bigl(1-\frac{1}{N^2}\bigr)\frac{1}{\pi^2\beta_0}\int p^4 \frac{g^2(\mu)}{1+\beta_0 g^2(\mu)\logp} e^{ip\cdot x}\frac{d^4p}{(2\pi)^4}\nonumber\\
&= \bigl(1-\frac{1}{N^2}\bigr)\frac{1}{\pi^2 \beta_0}\sum_{l=0}^{\infty}(-1)^l\int p^4 g^2(\mu)\bigl(\beta_0 g^2(\mu)\logp\bigr)^l e^{ip\cdot x}\frac{d^4p}{(2\pi)^4}
\end{align}
We extract the leading logarithms of this Fourier transform.
By leading we mean terms that have the highest power of logarithm with the power of $g$ fixed.
We use  Eq.(\ref{eqn:ft_leading}) that furnishes the leading logarithm of the Fourier transform:
\begin{equation}
\int p^4\biggl(\log\frac{p^2}{\mu^2}\biggr)^l e^{ip \cdot x}  \frac{d^4p}{(2\pi)^4} =
-\frac{l\Gamma(4) 2^{5}}{\pi^2}
\frac{1}{x^8}\biggl(\log\frac{4}{x^2\mu^2}\biggr)^{l-1}+\cdots
\end{equation}
Inserting it in Eq.(\ref{eqn:serie_p_da_trasformare}) we obtain for the leading logarithms: 
\begin{align}\label{eqn:x_series_da_tf}
& \bigl(1-\frac{1}{N^2}\bigr)\frac{1}{\pi^2 \beta_0}\int p^4 \frac{g^2(\mu)}{1+\beta_0 g^2(\mu)\logp} e^{ip\cdot x}\frac{d^4p}{(2\pi)^4}\nonumber\\
&= \bigl(1-\frac{1}{N^2}\bigr)\frac{1}{\pi^2 \beta_0}\sum_{l=0}^{\infty}(-1)^{l-1}g^2(\mu)\beta_0^l g^{2l}(\mu)\frac{l\cdot\Gamma(4) 2^5}{\pi^2}\frac{1}{x^8}\biggl(\log\frac{4}{x^2\mu^2}\biggr)^{l-1}
\end{align}  
We compare it with the $ASD$ correlator in the coordinate representation Eq.(\ref{eqn:corr_asd_{a}}):
\begin{align}\label{eqn:rg_improved_2l_{a}_series}
&\bigl(1-\frac{1}{N^2}\bigr)\frac{1}{\pi^2 \beta_0}\frac{ 2^5}{\pi^2 x^8}\Gamma(4) g^4(x)\sim\bigl(1-\frac{1}{N^2}\bigr)\frac{1}{\pi^2 \beta_0}
\frac{ 2^5}{\pi^2 x^8}\Gamma(4) \biggl(\frac{g^2(\mu)}{1+\beta_0 g^2(\mu)\log\frac{4}{x^2\mu^2}}\biggr)^2\nonumber\\
&=\bigl(1-\frac{1}{N^2}\bigr)\frac{1}{\pi^2 \beta_0}\frac{ 2^5}{\pi^2 x^8}\Gamma(4)g^4(\mu)\sum_{n=0}^{\infty}\sum_{l=0}^{\infty}(-1)^{n+l}\bigl(\beta_0 g^2(\mu)\log\frac{x^2\mu^2}{4}\bigr)^{n+l}
\end{align} 
We want to prove that the two series in Eq.(\ref{eqn:x_series_da_tf}) and in Eq.(\ref{eqn:rg_improved_2l_{a}_series}) are equal. 
The proof is by induction.
We prove it for the first non trivial term, i.e. for $l=1$ in Eq.(\ref{eqn:x_series_da_tf}):
\begin{equation}
\bigl(1-\frac{1}{N^2}\bigr)\frac{1}{\pi^2 \beta_0}
\frac{ 2^5}{\pi^2 x^8}\Gamma(4)
g^{4}(\mu)
\end{equation} 
that is equal to the term obtained from Eq.(\ref{eqn:rg_improved_2l_{a}_series}) putting $n=l=0$.
Assuming that the equality is valid up to the order of $\biggl(\log\frac{x^2\mu^2}{4} \biggr)^{m-1}$, we show that it holds at the order of $\biggl(\log\frac{x^2\mu^2}{4} \biggr)^{m}$.
Indeed, the $m$-power of the logarithm occurs in Eq.(\ref{eqn:x_series_da_tf}) for $l=m+1$:
\begin{align}\label{eqn:dim_induzione_intermedio}
&\bigl(1-\frac{1}{N^2}\bigr)\frac{1}{\pi^2 \beta_0}
\int p^4 \frac{g^2(\mu)}{1+\beta_0 g^2(\mu)\logp} e^{ip\cdot x}\frac{d^4p}{(2\pi)^4} \nonumber\\
&\sim \bigl(1-\frac{1}{N^2}\bigr)\frac{1}{\pi^2 \beta_0} \bigg[
\sum_{l=0}^{m}(-1)^{l-1}g^2(\mu)\beta_0^l g^{2l}(\mu)\frac{l\cdot\Gamma(4) 2^5}{\pi^2}\frac{1}{x^8}\biggl(\log\frac{4}{x^2\mu^2}\biggr)^{l-1}+\nonumber\\
&+(-1)^m\frac{(m+1)\cdot 2^5}{\pi^2 x^8}\beta_0^{m} \Gamma(4)
g^{2m+4}(\mu)\biggl(\log\frac{4}{x^2\mu^2}\biggr)^{m} \bigg]
\end{align} 
The $m$-th power of the logarithm in Eq.(\ref{eqn:rg_improved_2l_{a}_series}) occurs for the $m+1$ couples $(n,l)$ such that $l+n=m$:
\begin{align}
&\bigl(1-\frac{1}{N^2}\bigr)\frac{1}{\pi^2 \beta_0}
\frac{ 2^5}{\pi^2 x^8}\Gamma(4)g^4(x)\nonumber\\
&\sim \bigl(1-\frac{1}{N^2}\bigr)\frac{1}{\pi^2 \beta_0} \bigg[
\frac{ 2^5}{\pi^2 x^8}\Gamma(4)g^4(\mu)\sum_{n,l=0}^{\substack{\\n+l\leq m-1}}(-1)^{n+l}\bigl(\beta_0 g^2(\mu)\log\frac{4}{x^2\mu^2}\bigr)^{n+l}+\nonumber\\
&+\frac{ 2^5}{\pi^2 x^8}\Gamma(4)g^4(\mu)\sum_{n,l=0}^{\substack{\\n+l=m}}(-1)^{m}\bigl(\beta_0 g^2(\mu)\log\frac{4}{x^2\mu^2}\bigr)^{m} \bigg]\nonumber\\
&=\bigl(1-\frac{1}{N^2}\bigr)\frac{1}{\pi^2 \beta_0}\bigg[
\frac{ 2^5}{\pi^2 x^8}\Gamma(4)g^4(\mu)\sum_{n,l=0}^{\substack{\\n+l\leq m-1}}(-1)^{n+l}\bigl(\beta_0 g^2(\mu)\log\frac{4}{x^2\mu^2}\bigr)^{n+l}+\nonumber\\
&+(-1)^m\frac{(m+1)\cdot 2^5}{\pi^2 x^8}\beta_0^{m} \Gamma(4)
g^{2m+4}(\mu)\biggl(\log\frac{4}{x^2\mu^2}\biggr)^{m} \bigg]
\end{align}
For the inductive hypothesis the first term in the last expression is equal to the first one in Eq.(\ref{eqn:dim_induzione_intermedio}), i.e.:
\begin{align}
&\bigl(1-\frac{1}{N^2}\bigr)\frac{1}{\pi^2 \beta_0} \bigg[
\frac{ 2^5}{\pi^2 x^8}\Gamma(4)g^4(\mu)\sum_{n,l=0}^{\substack{\\n+l\leq m-1}}(-1)^{n+l}\bigl(\beta_0 g^2(\mu)\log\frac{x^2\mu^2}{4}\bigr)^{n+l}  \bigg]\nonumber\\
&=\bigl(1-\frac{1}{N^2}\bigr)\frac{1}{\pi^2 \beta_0} \bigg[
\sum_{l=0}^{m}(-1)^{l-1}g^2(\mu)\beta_0^l g^{2l}(\mu)\frac{l\cdot\Gamma(4) 2^5}{\pi^2}\frac{1}{x^8}\biggl(\log\frac{4}{x^2\mu^2}\biggr)^{l-1} \bigg]
\end{align}
The remaining terms, i.e. the terms of order of $\log^m\frac{4}{x^2\mu^2}$, are equal and therefore
the proof by induction is complete.

\section{ $ASD$ correlator in the Topological Field Theory}\thispagestyle{empty} 

We briefly summarize the results for the glueball propagators in the $TFT$ underlying large-$N$ $YM$ \cite{boch:quasi_pbs} \cite{MB0} \cite{boch:crit_points} \cite{boch:glueball_prop} \cite{Top}. 
For the $ASD$ glueball propagator \cite{boch:glueball_prop}\cite{boch:crit_points} \footnote{We use here a manifestly covariant notation
as opposed to the one in the $TFT$ \cite{boch:glueball_prop}\cite{boch:crit_points}.}:
\begin{equation}\label{eqn:intro_formula_L2}
\frac{1}{2}\braket{\frac{g^2}{N}tr\bigl(F^{-2}(p)\bigr) \frac{g^2}{N}tr \bigl(F^{-2}(-p)\bigr)}_{conn} =
\frac{1}{\pi^2}\sum_{k=1}^{\infty}\frac{k^2 g_k^4\Lambda_{\overline{W}}^6 }{p^2+k\Lambda_{\overline{W}}^2} + ...
\end{equation}
Besides, in the $TFT$ the two-point correlators of certain scalar operators $\mathcal{O}_{2L}$ of naive dimension $D=2L$ that are homogeneous polynomials of degree $L$
in the $ASD$ curvature $F^-$\cite{boch:glueball_prop}\cite{boch:crit_points} can be computed asymptotically for large $L$:
\begin{align}\label{eqn:formula}
&\braket{\mathcal{O}_{2L}(p)\mathcal{O}_{2L}(-p)}_{conn}
=const\sum_{k=1}^{\infty}\frac{k^{2L-2} Z_k^{-L}\Lambda_{\overline{W}}^2 \Lambda_{\overline{W}}^{4L-4}}{p^2+k\Lambda_{\overline{W}}^2} 
\end{align}
The operators $\mathcal{O}_{2L}$ occur as the ground state in the integrable sector of large-$N$ $YM$ of Ferreti-Heise-Zarembo \cite{ferretti:new_struct} asymptotically for large $L$.
Ferreti-Heise-Zarembo have computed their one-loop anomalous dimension for large $L$ \cite{ferretti:new_struct}:
\begin{align}
\gamma_{0 (\mathcal{O}_{2L})}= \frac{1}{(4\pi)^2}\frac{5}{3} L+O(\frac{1}{L})
\end{align}
The ground state for $L=2$ is the $ASD$ operator that occurs in Eq.(\ref{eqn:intro_formula_L2}) for which $\gamma_{0 (\mathcal{O}_{4})}=2 \beta_0$ exactly. \par
In Eq.(\ref{eqn:intro_formula_L2}) and in Eq.(\ref{eqn:formula}) $\Lambda_{\overline{W}}$ is the $RG$ invariant scale in the scheme in which it coincides with the mass gap. The functions $g^2(\frac{p^2}{\Lambdawb^2})$ and $Z(\frac{p^2}{\Lambdawb^2})$ are the solutions of the differential equations:
\begin{align} \label{eqn:eq_def_gk}
\frac{\partial g}{\partial \log p}
&=\frac{-\beta_0 g^3+\frac{1}{(4\pi)^2}g^3\frac{\partial \log Z}{\partial \log p}}{1-\frac{4}{(4\pi)^2}g^2} \nonumber \\
\frac{\partial\log Z}{\partial\log p}
&=2\gamma_0 g^2 +\cdots \nonumber \\
\gamma_{0}&=\frac{1}{(4\pi)^2}\frac{5}{3}
\end{align}
where $p$ is equal to the square root of $p^2$.
The definitions of $g_k$ and $Z_k$ are:
\begin{align}
&g_k=g(k)\\
&Z_k=Z(k)
\end{align}
In \cite{boch:quasi_pbs} it is shown that Eq.(\ref{eqn:eq_def_gk}) reproduces the correct universal one-loop and two-loop coefficients of the perturbative $\beta$ function of pure $YM$. Indeed, substituting in Eq.(\ref{eqn:eq_def_gk}) we get:
\begin{align}\label{eqn:matching_beta_pert}
\frac{\partial g}{\partial \log p}
&=\frac{-\beta_0 g^3+\frac{2\gamma_0 }{(4\pi)^2}g^5}{1-\frac{4}{(4\pi)^2}g^2}+\cdots \nonumber \\
&=\bigl(-\beta_0 g^3+\frac{2\gamma_0}{(4\pi)^2}g^5\bigr)\bigl(1+\frac{4}{(4\pi)^2}g^2\bigr)+\cdots \nonumber\\
&=-\beta_0 g^3+\frac{2\gamma_0}{(4\pi)^2}  g^5-\frac{4\beta_0}{(4\pi)^2}g^5+\cdots \nonumber\\
&=-\beta_0 g^3+\frac{1}{(4\pi)^4}\frac{10}{3}g^5 - \frac{44}{3}\frac{1}{(4\pi)^4}g^5+\cdots\nonumber\\
&=-\beta_0 g^3-\beta_1 g^5+\cdots
\end{align}
where:
\begin{align}
&\beta_0=\frac{1}{(4\pi)^2}\frac{11}{3}\\
&\beta_1=\frac{1}{(4\pi)^4}\frac{34}{3}
\end{align}
These are the correct one- and two-loop coefficients that arise in perturbation theory of pure $YM$ for the 't Hooft coupling.
Therefore, the renormalization-group improved universal asymptotic behavior of $g_k$ is:
\begin{equation}\label{eqn:gk_as_behav}
g^2_k\sim\frac{1}{\beta_0\log\frac{k}{c}}\biggl(1-\frac{\beta_1}{\beta_0^2}\frac{\log\log\frac{k}{c}}{\log\frac{k}{c}}\biggr)
+O\biggl(\frac{1}{\log^2\frac{k}{c}}\biggr)
\end{equation}
and the renormalization group improved universal asymptotic behavior of $Z_k^{-1}$ is:
\begin{equation}\label{eqn:zk_as_behav}
Z_k^{-1}\sim (g^2_k)^{\frac{\gamma_0}{\beta_0}}\sim
\Biggl(\frac{1}{\beta_0\log\frac{k}{c}}\biggl(1-\frac{\beta_1}{\beta_0^2}\frac{\log\log\frac{k}{c}}{\log\frac{k}{c}}\biggr)+O\biggl(\frac{1}{\log^2\frac{k}{c}} \biggr)  \Biggr)^{\frac{\gamma_0}{\beta_0}}
\end{equation}
In this section we find the asymptotics of the $ASD$ propagator in Eqs.(\ref{eqn:intro_formula_L2}) and of the large-$L$ propagator in Eq.(\ref{eqn:formula}) at the order of the leading and of the next-to-leading logarithms following the technique employed in \cite{boch:glueball_prop} at the order of the leading logarithm. \par
To find the asymptotics of the glueball propagator for large $L$ in Eq.(\ref{eqn:formula}) we follow the strategy explained in sect.(1.4) for the
$ASD$ correlator. Firstly, we highlight the physical terms contained in Eq.(\ref{eqn:formula}) neglecting the non-physical contact terms. Secondly, we extract the asymptotic behavior writing the sum in Eq.(\ref{eqn:formula}) as an integral \cite{boch:glueball_prop}. Finally, we use the leading and next-to-leading expression for $Z_k^{-1}$ in Eq.(\ref{eqn:zk_as_behav}) to compare Eq.(\ref{eqn:formula}) with $RG$-improved perturbation theory.

We write Eq.(\ref{eqn:formula}) as \cite{boch:glueball_prop}\cite{boch:crit_points}:
\begin{align}
&\sum_{k=1}^{\infty}
\frac{k^{2(L-1)}Z_k^{-L}\Lambda_{\overline{W}}^2 \Lambda_{\overline{W}}^{4(L-1)}}{p^2+k\Lambda_{\overline{W}}^2}\nonumber\\
&=\sum_{k=1}^{\infty}
\frac{((k\Lambda_{\overline{W}}^2+p^2)(k\Lambda_{\overline{W}}^2-p^2)+p^4)^{L-1} Z_k^{-L}\Lambda_{\overline{W}}^2}{p^2+k\Lambda_{\overline{W}}^2}\nonumber\\
&=p^{4L-4}\sum_{k=1}^{\infty}\frac{Z_k^{-L}\Lambda_{\overline{W}}^2 }{p^2+k\Lambda_{\overline{W}}^2} \nonumber\\
&+\sum_{k=1}^{\infty}\sum_{m=1}^{L-1}\binom{L-1}{m}p^{4(L-1-m)}(k\Lambda_{\overline{W}}^2+p^2)^{m-1}(k\Lambda_{\overline{W}}^2-p^2)^m  Z_k^{-L}\Lambda_{\overline{W}}^2\nonumber\\ 
&\sim p^{4L-4}\sum_{k=1}^{\infty}\frac{ Z_k^{-L}\Lambda_{\overline{W}}^2 }{p^2+k\Lambda_{\overline{W}}^2}+\dots
\end{align}
where the dots stand for contact terms.

As in sect.(1.4) we use the Euler-McLaurin formula to approximate the sum to an integral \cite{boch:glueball_prop}\cite{boch:crit_points}:
\begin{equation}
\sum_{k=k_1}^{\infty}G_k(p)=
\int_{k_1}^{\infty}G_k(p)dk - \sum_{j=1}^{\infty}\frac{B_j}{j!} \left[\partial_k^{j-1}G_k(p)\right]_{k=k_1}
\end{equation}
In our case the terms proportional to the Bernoulli numbers involve negative powers of $p$ and they are therefore subleading with respect to the first term, hence we ignore them. 

We obtain: 
\begin{equation}\label{eqn:int_fond}
\sum_{k=1}^{\infty}\frac{Z_k^{-L}\Lambda_{\overline{W}}^2 }{p^2+k\Lambda_{\overline{W}}^2}\sim
\int_{1}^\infty\frac{Z_k^{-L}}{k+\frac{p^2}{\Lambda_{\overline{W}}^2}}dk
\end{equation}
In order to compare Eq.(\ref{eqn:int_fond}) to the $RG$-improved perturbation theory, we substitute for $Z_k^{-1}$ its leading and next-to-leading logarithmic behavior given by Eq.(\ref{eqn:zk_as_behav}).
We define:
\begin{equation}
\gamma'=\frac{\gamma_0}{\beta_0}L
\end{equation}
and:
\begin{equation}
\nu=\frac{p^2}{\Lambda_{\overline{W}}^2}
\end{equation}
The integral that determines the leading asymptotic behavior is:
\begin{equation}\label{eqn:int_fond2}
I_c^{1}(\nu)=\int_{1}^\infty\biggl(\frac{1}{\beta_0\log(\frac{k}{c})}\biggr)^{\gamma'}\frac{dk}{k+\nu}
\end{equation}
The next-to-leading logarithmic behavior is determined by:
\begin{equation}\label{eqn:int_fond3}
I^{2}_c(\nu)=\int_{1}^\infty\left(\frac{1}{\beta_0\log(\frac{k}{c})}\left(1-\frac{\beta_1}{\beta_0^2}\frac{\log\log(\frac{k}{c})}{\log(\frac{k}{c})}\right)\right)^{\gamma'}\frac{dk}{k+\nu}
\end{equation}
$\gamma'=2$ for the $ASD$ correlator and $\gamma'=\frac{\gamma_0}{\beta_0}L$ for the large-$L$ correlator.
We show in the following that the leading and next-to-leading behavior of $I^{2}_c(\nu)$ is: 
\begin{align}
&I^{2}_c(\nu)\sim\frac{1}{\gamma_0 L-\beta_0}\Biggl[\frac{1}{\beta_0\log\plw}\biggl(1-\frac{\beta_1}{\beta_0^2}\frac{\log\logplw}{\logplw}\biggr)\Biggr]^{\frac{\gamma_0}{\beta_0}L-1} 
\end{align}
Therefore, the asymptotic behavior of the correlator of the $TFT$ for large-$L$ is:
\begin{equation}
\braket{\mathcal{O}_{2L}(p)\mathcal{O}_{2L}(-p)}_{conn}\sim p^{4L-4}\frac{1}{\gamma_0 L-\beta_0}\bigl(g^2(p)\bigr)^{\frac{\gamma_0}{\beta_0}L-1}
\end{equation}
It agrees with the naive $RG$ estimate Eq.(\ref{eqn:naive_rg}).

\subsection{Asymptotic series to the order of the leading logarithm}

We now perform an explicit expansion in series of $I_c^{1}(\nu)$.
Firstly, we change variables from $k$ to $k+\nu$:
\begin{equation}
I_c^{1}(\nu)=\int_{1+\nu}^\infty\biggl(\frac{1}{\beta_0\log(\frac{k-\nu}{c})}\biggr)^{\gamma'}\frac{dk}{k}
\end{equation}
We have that: 
\begin{equation}
[\log(\frac{k'-\nu}{c})]^{-\gamma'}=
[\log(\frac{k'}{c})]^{-\gamma'}\biggl[1+\frac{\log(1-\frac{\nu}{k'})}{\log(\frac{k'}{c})}\biggr]^{-\gamma'}
\end{equation}
It is easy to see that if $c<1$:
\begin{equation}
\frac{\log(1-\frac{\nu}{k'})}{\log(\frac{k'}{c})}<1
\end{equation}
We define:
\begin{equation}
\epsilon=\frac{\log(1-\frac{\nu}{k'})}{\log(\frac{k'}{c})}
\end{equation}
and we exploit the binomial formula \cite{BIN}:
\begin{align}\label{eqn:espansione_bin}
(1+\epsilon)^{-\gamma'}&=\sum_{r=0}^{\infty}\binom{\gamma'+r-1}{r}(-1)^r\epsilon^r
\end{align}
to obtain a series expansion.
We proceed order by order in $\epsilon$.
At the order of $\epsilon^1$ the only contribution is:
\begin{equation}
-\gamma'\epsilon
\end{equation}
$\epsilon$ can be further expanded in powers of $\eta=\frac{\nu}{k'}$, since in the integration domain $\eta<1$:
\begin{align}\label{eqn:espansione_log}
\log(1-\eta)&=\sum_{m=1}^{\infty}\frac{(-1)^{2m+1}}{m}\eta^m
\end{align}
Up to the order of $\eta^1$ this expansion reads: 
\begin{equation}
-\gamma'\epsilon\sim\gamma'\frac{\nu}{k'\log(\frac{k'}{c})} 
\end{equation} 
Substituting in $I_c^{1}(\nu)$ we get:
\begin{align}\label{eqn: int1}
&\int_{1+\nu}^\infty\frac{1}{k'}[\beta_0\log(\frac{k'}{c})]^{-\gamma'}
\biggl[1+\frac{\log(1-\frac{\nu}{k'})}{\log(\frac{k'}{c})}\biggr]^{-\gamma'}dk'\nonumber\\
&\sim \int_{1+\nu}^\infty\frac{1}{k'}[\beta_0\log(\frac{k'}{c})]^{-\gamma'} 
\biggl[1+\gamma'\frac{\nu}{k'\log(\frac{k'}{c})}\biggr]dk'\nonumber\\
&=\int_{1+\nu}^\infty\frac{1}{k'}[\beta_0\log(\frac{k'}{c})]^{-\gamma'}dk'+
\gamma' \nu\int_{1+\nu}^\infty\frac{1}{k'^2}\beta_0^{-\gamma'}[\log(\frac{k'}{c})]^{-\gamma'-1}dk' 
\end{align}
From the first integral it follows the leading asymptotic behavior \cite{boch:glueball_prop}:
\begin{equation}\label{eqn:sol_int_leading}
\int_{1+\nu}^\infty\frac{1}{k'}[\beta_0\log(\frac{k'}{c})]^{-\gamma'}dk'=
\frac{1}{\gamma'-1} \beta_0^{-\gamma'}\left[\log\left(\frac{1+\nu}{c}\right)\right]^{-\gamma'+1}
\end{equation}
Since for large $\nu$:
\begin{equation}
\biggl(\log\left(\frac{1+\nu}{c}\right)\biggr)^{-1}\sim
\bigl(\log\nu\bigr)^{-1} 
\end{equation}
it follows the leading asymptotic behavior of Eq.(\ref{eqn:formula}) \cite{boch:glueball_prop}:
\begin{align}
\braket{\mathcal{O}_{2L}(p)\mathcal{O}_{2L}(-p)}_{conn}\sim\frac{p^{4L-4}}{\gamma_0 L-\beta_0}\Biggl[\frac{1}{\beta_0\log\plw}\Biggr]^{\frac{\gamma_0}{\beta_0}L-1}
\end{align} 
Performing the same steps for the $ASD$ correlator, i.e. for $L=2$ and $\gamma'=2$, we get:
\begin{equation}
\frac{1}{\pi^2}\int_1^{\infty}\frac{\bigl(\beta_0\log\frac{k}{c}\bigr)^{-2}}{k+\nu}\sim
\frac{1}{\pi^2\beta_0}\Biggl(\beta_0\log\biggl(\frac{1+\frac{p^2}{\Lambda_{\overline{W}}^2}}{c}\biggr)\Biggr)^{-1}\sim
\frac{1}{\pi^2\beta_0}\frac{1}{\beta_0\logplw}
\end{equation}
that agrees with the leading logarithm of the asymptotic behavior in Eq.(\ref{eqn:corr_asd}). \par
We now compute the second term in the last line of Eq.(\ref{eqn: int1}), that is the first subleading term. 
We write it as:
\begin{equation}\label{eqn:int_ord1}
\frac{\gamma'\nu \beta_0^{-\gamma'}}{c}\int_{\frac{1+\nu}{c}}^\infty\frac{1}{k^2}[\log(k)]^{-\gamma'-1}dk 
\end{equation}
and we integrate by parts:
\begin{align}
&\frac{\gamma'\nu \beta_0^{-\gamma'}}{c}\int_{\frac{1+\nu}{c}}^{\infty}\frac{1}{k^2}[\log(k)]^{-\gamma'-1}dk \nonumber\\
&=\frac{\gamma'\nu \beta_0^{-\gamma'}}{c}\biggl[-\frac{[\log(k)]^{-\gamma'-1}}{k}\bigg|_{\frac{1+\nu}{c}}^{\infty}
-(\gamma'+1)\int_{\frac{1+\nu}{c}}^{\infty}\frac{dk}{k^2}[\log(k)]^{-\gamma'-2}\biggr]\nonumber\\
&=\frac{\gamma'\nu \beta_0^{-\gamma'}}{c}\biggl[\frac{c}{1+\nu}[\log(\frac{1+\nu}{c})]^{-\gamma'-1}
-(\gamma'+1)\int_{\frac{1+\nu}{c}}^{\infty}\frac{dk}{k^2}[\log(k)]^{-\gamma'-2}\biggr]
\end{align}
We notice that the second term in the last line has the same structure as the original integral but with a more negative power of the logarithm.
This implies that it is a less relevant term.
Furthermore, since performing integration by parts repeatedly we always obtain integrals with the same structure, we can derive a possibly asymptotic series expansion for Eq.(\ref{eqn:int_ord1}): 
\begin{align}
&\frac{\gamma'\nu \beta_0^{-\gamma'}}{c}\int_{\frac{1+\nu}{c}}^{\infty}\frac{dk}{k^2}[\log(k)]^{-\gamma'-1}\nonumber\\
&=\beta_0^{-\gamma'}\frac{\nu}{1+\nu}\sum_{s=0}^{\infty}(-1)^s\left(\prod_{t=0}^{s}(\gamma'+t)\right)[\log(\frac{1+\nu}{c})]^{-\gamma'-1-s}\nonumber\\
&=\beta_0^{-\gamma'}\frac{p^2}{p^2+\Lambda_{\overline{W}}^2}\sum_{s=0}^{\infty}(-1)^s\left(\prod_{t=0}^{s}(\gamma'+t)\right)
\biggl[\log\biggl(\frac{1+\frac{p^2}{\Lambda_{\overline{W}}^2}}{c}\biggr)\biggr]^{-\gamma'-1-s}
\end{align}

Now that we have understood the technique, we derive a complete expression taking into account all the terms coming from the expansion of the logarithm in Eq.(\ref{eqn:espansione_log}),
simply substituting it in $I_c^{1}(\nu)$: 
\begin{align}
&\int_{1+\nu}^{\infty}\frac{dk'}{k'}[\beta_0\log(\frac{k'}{c})]^{-\gamma'}
\biggl[1-\gamma'\frac{\sum_{m=1}^{\infty}\frac{(-1)^{2m+1}\nu^m}{m k^m}}{\log(\frac{k'}{c})}\biggr]\nonumber\\
&=\int_{1+\nu}^{\infty}\frac{dk'}{k'}[\beta_0\log(\frac{k'}{c})]^{-\gamma'}\nonumber\\
&-\gamma'\beta_0^{-\gamma'}
\sum_{m=1}^{\infty}\frac{(-1)^{2m+1}\nu^m}{m}\int_{1+\nu}^{\infty}\frac{dk'}{k'^{m+1}}\left[\log(\frac{k'}{c})\right]^{-\gamma'-1}
\end{align}
Focusing on the second term:
\begin{align}\label{eqn:ord1_eps}
&\gamma'\beta_0^{-\gamma'}
\sum_{m=1}^{\infty}\frac{(-1)^{2m+1}\nu^m}{m}\int_{1+\nu}^{\infty}\frac{dk'}{k'^{m+1}}\left[\log(\frac{k'}{c})\right]^{-\gamma'-1}\nonumber\\
&=\beta_0^{-\gamma'}\gamma'\sum_{m=1}^{\infty}\frac{(-1)\nu^m}{m c^m}\left[-\frac{\left[\log(k)\right]^{-\gamma'-1}}{m k^{m}}
\bigg|_{\frac{1+\nu}{c}}^{\infty}-
(\gamma'+1)\int_{\frac{1+\nu}{c}}^{\infty}dk\frac{\left[\log(k)\right]^{-\gamma'-2}}{mk^{m+1}}\right]\nonumber\\
&=\beta_0^{-\gamma'}\sum_{m=1}^{\infty}\frac{(-1)\nu^m}{m c^m}\left[\sum_{s=0}^{\infty}\frac{(-1)^{s}}{m^{s+1}}
\frac{c^m\prod_{t=0}^{s}(\gamma'+t)}{(1+\nu)^m}\left[\log\left(\frac{1+\nu}{c}\right)\right]^{-\gamma'-1-s}\right]\nonumber\\
&=\beta_0^{-\gamma'}\sum_{m=1}^{\infty}\sum_{s=0}^{\infty}(-1)^{s+1}\left(\frac{\nu}{1+\nu}\right)^m\frac{\prod_{t=0}^{s}(\gamma'+t)}{m^{s+2}}
\left[\log\left(\frac{1+\nu}{c}\right)\right]^{-\gamma'-s-1}	
\end{align}
Therefore, at the first order in $\epsilon$ we get:
\begin{align}\label{eqn:int_lead_temp}
&\int_{1+\nu}^\infty dk'\frac{1}{k'}[\beta_0\log(\frac{k'}{c})]^{-\gamma'}
\biggl[1+\frac{\log(1-\frac{\nu}{k'})}{\log(\frac{k'}{c})}\biggr]^{-\gamma'}\nonumber\\
&\sim \int_{1+\nu}^{\infty}\frac{dk'}{k'}[\beta_0\log(\frac{k'}{c})]^{-\gamma'} \nonumber\\
&+\beta_0^{-\gamma'}\sum_{m=1}^{\infty}\sum_{s=0}^{\infty}(-1)^{s}\left(\frac{\nu}{1+\nu}\right)^m\frac{\prod_{t=0}^{s}(\gamma'+t)}{m^{s+2}}
\left[\log\left(\frac{1+\nu}{c}\right)\right]^{-\gamma'-s-1}	 
\end{align}
We find the subleading behavior keeping only the terms with $s=0$ in Eq.(\ref{eqn:int_lead_temp}).
We obtain in the large $\nu$ limit: 
 \begin{align}\label{eqn:correzione_serie_as}
&\int_{1+\nu}^\infty dk'\frac{1}{k'}[\beta_0\log(\frac{k'}{c})]^{-\gamma'}
\biggl[1+\frac{\log(1-\frac{\nu}{k'})}{\log(\frac{k'}{c})}\biggr]^{-\gamma'}\nonumber\\
& \sim \int_{1+\nu}^{\infty}\frac{dk'}{k'}[\beta_0\log(\frac{k'}{c})]^{-\gamma'}
+\gamma'\beta_0^{-\gamma'}\left[\log \nu \right]^{-\gamma'-1}\sum_{m=0}^{\infty}\frac{1}{m^2}\nonumber\\
& \sim \frac{1}{\gamma'-1}\beta_0^{-\gamma'}\left[\log\nu\right]^{-\gamma'+1}
+\gamma'\beta_0^{-\gamma'}\zeta(2)[\log \nu]^{-\gamma'-1}
\end{align} 
It is interesting to notice that the transcendental function $\zeta(2)=\frac{\pi^2}{6}$ occurs, as it often does in Feynman-graph computations.

\subsection{Asymptotic series to the order of the next-to-leading logarithm}

We now perform a series expansion of $I_c^{2}(\nu)$:
\begin{align}
I_c^{2}(\nu)&=\int_1^{\infty}\beta_0^{-\gamma'}\left(\frac{1}{\log(\frac{k}{c})}\left(1-\frac{\beta_1}{\beta_0^2}\frac{\log\log(\frac{k}{c})}{\log(\frac{k}{c})}\right)\right)^{\gamma'}\frac{dk}{k+\nu}\nonumber\\
&=\beta_0^{-\gamma'}\int_{1+\nu}^{\infty}\left(\frac{1}{\log(\frac{k-\nu}{c})}\left(1-\frac{\beta_1}{\beta_0^2}\frac{\log\log(\frac{k-\nu}{c})}{\log(\frac{k-\nu}{c})}\right)\right)^{\gamma'}\frac{dk}{k}\nonumber\\
&\sim \beta_0^{-\gamma'}\int_{1+\nu}^{\infty}\left[\log(\frac{k-\nu}{c})\right]^{-\gamma'}
\left(1-\gamma'\frac{\beta_1}{\beta_0^2}\frac{\log\log(\frac{k-\nu}{c})}{\log(\frac{k-\nu}{c})}\right)\frac{dk}{k}\nonumber\\
&\sim  \beta_0^{-\gamma'}\int_{1+\nu}^{\infty}\left[\log(\frac{k-\nu}{c})\right]^{-\gamma'}\frac{dk}{k}+\nonumber\\
&-\gamma'\frac{\beta_1}{\beta_0^2}\beta_0^{-\gamma'}\int_{1+\nu}^{\infty}\left[\log(\frac{k-\nu}{c})\right]^{-\gamma'-1}\log\log(\frac{k-\nu}{c})
\frac{dk}{k}
\end{align}
The first integral has been evaluated in the previous section and the second term is the new contribution. We evaluate it at the leading order by changing variables and integrating by parts:
\begin{align}\label{eqn:int_next-to-leading}
&\gamma'\frac{\beta_1}{\beta_0^2}\beta_0^{-\gamma'}\int_{1+\nu}^{\infty}\left[\log(\frac{k-\nu}{c})\right]^{-\gamma'-1}\log\log(\frac{k-\nu}{c})
\frac{dk}{k}\nonumber\\
&\sim  \gamma'\frac{\beta_1}{\beta_0^2}\beta_0^{-\gamma'}\int_{1+\nu}^{\infty}\left[\log(\frac{k}{c})\right]^{-\gamma'-1}\log\log(\frac{k}{c})
\frac{dk}{k}\nonumber\\
&=\gamma'\frac{\beta_1}{\beta_0^2}\beta_0^{-\gamma'}\int_{\log\frac{1+\nu}{c}}^{\infty}t^{-\gamma'-1}\log(t)dt\nonumber\\
&=\gamma'\frac{\beta_1}{\beta_0^2}\beta_0^{-\gamma'}\left[\frac{1}{\gamma'}\left(\log(\frac{1+\nu}{c})\right)^{-\gamma'}\log\log(\frac{1+\nu}{c})+
\frac{1}{\gamma'^2}\left(\log(\frac{1+\nu}{c})\right)^{-\gamma'}\right]
\end{align}
The second term in brackets is subleading with respect to the first one. 
Putting together  Eq.(\ref{eqn:int_next-to-leading}) and Eq.(\ref{eqn:sol_int_leading}) we get for $I_c^{2}(\nu)$ :
\begin{align}\label{eqn:esp_ntl}
&\beta_0^{-\gamma'}\int_1^{\infty}\left(\frac{1}{\log(\frac{k}{c})}\left(1-\frac{\beta_1}{\beta_0^2}\frac{\log\log(\frac{k}{c})}{\log(\frac{k}{c})}\right)\right)^{\gamma'}\frac{dk}{k+\nu}\nonumber\\
& \sim \frac{1}{\gamma'-1}\beta_0^{-\gamma'}\left(\log\frac{1+\nu}{c}\right)^{-\gamma'+1}-\frac{\beta_1}{\beta_0^2}\beta_0^{-\gamma'}\left(\log(\frac{1+\nu}{c})\right)^{-\gamma'}\log\log(\frac{1+\nu}{c})\nonumber\\
&=\frac{\beta_0^{-\gamma'}}{\gamma'-1}\biggl(\log\frac{1+\nu}{c}\biggr)^{-\gamma'+1}\left[1-\frac{\beta_1(\gamma'-1)}{\beta_0^2}\left(\log(\frac{1+\nu}{c})\right)^{-1}\log\log(\frac{1+\nu}{c})\right]\nonumber\\
&\sim \frac{1}{\beta_0(\gamma'-1)}\left(\beta_0\log\frac{1+\nu}{c}\right)^{-\gamma'+1}\left[1-\frac{\beta_1}{\beta_0^2}\left(\log(\frac{1+\nu}{c})\right)^{-1}\log\log(\frac{1+\nu}{c})\right]^{\gamma'-1}\nonumber\\
&\sim \frac{1}{\beta_0(\gamma'-1)}(g^2(p))^{\gamma'-1}+ O\biggl(\Bigl(\frac{1}{\log\frac{p^2}{\Lambda_{\overline{W}}}}\Bigr)^{\gamma'}\biggr)
\end{align}
This result agrees with the $RGI$ perturbative estimate in Eq.(\ref{eqn:naive_rg}).
Repeating the same steps for the $ASD$ correlator we get: 
\begin{align}
&\frac{1}{\pi^2}\int_{1+\nu}^\infty\frac{1}{k'}[\beta_0\log(\frac{k'}{c})]^{-2}dk'
-\frac{\beta_1}{\pi^2\beta_0^4}\int_{1+\nu}^{\infty}\left[\log(\frac{k-\nu}{c})\right]^{-3}\log\log(\frac{k-\nu}{c})
\frac{dk}{k}\nonumber\\
& \sim \frac{1}{\pi^2\beta_0} g^2(p)+O\biggl(\frac{1}{\log^2\frac{p^2}{\Lambda_{\overline{W}}}}\biggr)
\end{align}
Again this result agrees with the universal behavior of the $RG$-improved perturbation theory in Eq.(\ref{eqn:rg_improved_scalar_2l}).

\newpage
\thispagestyle{empty}

\subsection{Link with the Lerch transcendent and the polylogarithmic function}

We may obtain the asymptotic behavior by a different method as an independent check, relating the relevant integrals
to special functions and employing the known asymptotic behavior of the special functions. \par

We briefly recall the definition of the Lerch Zeta function \cite{handbook, wikilerch}:
\begin{equation}
L(\lambda,s,a)=\sum_{n=0}^{\infty}\frac{e^{2\pi i\lambda n}}{(n+a)^s}
\end{equation}
Setting $z=e^{2\pi i \lambda}$, we obtain the Lerch transcendent \cite{handbook, wikilerch}:
\begin{equation}
\Phi(z,s,a)=\sum_{n=0}^{\infty}\frac{z^n}{(n+a)^s}
\end{equation}
The Lerch transcendent admits the integral representation:
\begin{equation}\label{eqn:lerch_int_repr}
\Phi(z,s,a)=\frac{1}{\Gamma(s)}\int_0^{\infty}\frac{t^{s-1}e^{-at}}{1-ze^{-t}}dt
\end{equation}
which is valid for $\Re (a)>0 \,\wedge \, \Re (s)>0 \,\wedge \,|z|<1$ or $\Re (a)>0 \,\wedge \, \Re (s)>1 \,\wedge \,|z|=1$.
The Lerch transcendent can be analytically continued to the region \cite{analytic}:
\begin{equation}
\mathcal{M}=\{(z,s,a)\in (\mathbb{C}\setminus \{0\})\times\mathbb{C}\times (\mathbb{C}\setminus\mathbb{Z})\}
\end{equation} 
Moreover, we exploit the following recursive formula:
\begin{equation}\label{eqn:lerch_recursive}
\Phi(z,s,a)=z^l \Phi(z,s,a+l)+\sum_{k=0}^{l-1}\frac{z^k}{(a+k)^s}
\end{equation}
Finally, we use the relationship between the Lerch transcendent and the polylogarithmic function \cite{handbook, wikipoly}:
\begin{equation}
\Li_s(z)=z\Phi(z,s,1)
\end{equation}
where the polylogarithmic function is defined by:
\begin{equation}
\Li_s(z)=\sum_{k=1}^{\infty}\frac{z^k}{k^s}
\end{equation}

\subsection{Asymptotic behavior and polylogarithmic function}

We start performing the change of variables $t=\log\frac{k}{c}$ in the integral in Eq.(\ref{eqn:int_fond2}):
\begin{equation}
I^1_c(\nu)=\int_{1}^\infty\frac{[\beta_0\log(\frac{k}{c})]^{-\gamma'}}{k+\nu}dk=
c\beta_0^{-\gamma'}\int_{\log\frac{1}{c}}^{\infty}\frac{t^{-\gamma'}}{c+\nu e^{-t}}dt
\end{equation}
Setting $c=e^{-\epsilon}$ in the limit $\epsilon\rightarrow 0$ we get the upper bound:
\begin{equation}
I^1_c(\nu)=\beta_0^{-\gamma'}\int_{\log\frac{1}{c}}^{\infty}\frac{t^{-\gamma'}}{1+\frac{\nu}{c} e^{-t}}dt\leq
\beta_0^{-\gamma'}\int_{\epsilon}^{\infty}\frac{t^{-\gamma'}}{1+\frac{\nu}{e^{-\epsilon}} e^{-t}}dt = I^1_{1-\epsilon}(\nu)
\end{equation}
but the upper bound is in fact asymptotic since varying $c$  is equivalent to a change of scheme.
Therefore, we take the limit $\epsilon\rightarrow 0$ in order to express $I^1_1$ in terms of the integral representation of the Lerch transcendent in Eq.(\ref{eqn:lerch_int_repr}). We get:
\begin{equation}
I^1_1(\nu)=\beta_0^{-\gamma'}\Gamma(-\gamma'+1)\Phi(-\nu,-\gamma'+1,0)
\end{equation}
We now exploit the relation in Eq.(\ref{eqn:lerch_recursive}) with
$n=1$, $a=0$, $z=-\nu$ and $s=-\gamma'+1$: 
\begin{equation}
\Phi(-\nu,-\gamma'+1,0)=z \Phi(-\nu,-\gamma'+1,1)
\end{equation}
Finally, we find the relation with the polylogarithmic function:
\begin{equation}
I^1_1(\nu)=\beta_0^{-\gamma'}\Gamma(-\gamma'+1)\Li_{-\gamma'+1}(-\nu)
\end{equation}
Now we use the following asymptotic expansion of $\Li_s$ \cite{wikipoly}:
\begin{equation}
\Li_s(z)=\sum_{j=0}^{\infty}(-1)^j(1-2^{1-2j})(2\pi)^{2j}\frac{B_{2j}}{(2j)!}\frac{[\log(-z)]^{s-2j}}{\Gamma(s+1-2k)}
\end{equation}
to find an asymptotic expansion for $I^1_1(\nu)$:
\begin{equation}\label{eqn:esp_polylog}
I^1_1(\nu)=\beta_0^{-\gamma'}\Gamma(-\gamma'+1)\sum_{j=0}^{\infty}(-1)^j(1-2^{1-2j})(2\pi)^{2j}\frac{B_{2j}}{(2j)!}
\frac{[\log\nu]^{-\gamma'+1-2j}}{\Gamma(-\gamma'+2-2j)}
\end{equation}
We get the leading behavior of $I_1^1(\nu)$ from the $j=0$ term in Eq.(\ref{eqn:esp_polylog}):
\begin{equation}\label{eqn:leading_polylog}
I_1^1(\nu)\sim -\beta_0^{-\gamma'}\Gamma(-\gamma'+1)\frac{[\log \nu]^{-\gamma'+1}}{\Gamma(-\gamma'+2)}=
\frac{[\beta_0\log \nu]^{-\gamma'+1}}{\beta_0(\gamma'-1)}
\end{equation}
Keeping also the $j=1$ term we obtain: 
\begin{equation}\label{eqn:subleading_polylog}
I_1^1(\nu)\sim \frac{[\beta_0\log \nu]^{-\gamma'+1}}{\beta_0(\gamma'-1)}+
\gamma'\beta_0^{-\gamma'}\frac{\pi^2}{6}[\log \nu]^{-\gamma'-1} 
\end{equation}
in perfect agreement with Eq.(\ref{eqn:correzione_serie_as}) since $\zeta(2)=\frac{\pi^2}{6}$.
Reinserting the momentum $p$ in the definition of $\nu$ the asymptotic result is:
\begin{equation}
I^1_c\left(\frac{p^2}{\Lambda_{\overline{W}}^2}\right)\sim \frac{[\beta_0\log(\frac{p^2}{\Lambda_{\overline{W}}^2})]^{-\frac{\gamma_0}{\beta_0}L+1}}{\gamma_0 L-\beta_0}+
\gamma_0 L\frac{\pi^2}{6}[\beta_0\log(\frac{p^2}{\Lambda_{\overline{W}}^2})]^{-\frac{\gamma_0}{\beta_0}L-1}  
\end{equation}
Using the same technique we find the next-to-leading logarithmic behavior of $I^2_c$. 
Indeed, also in this case we obtain an upper bound putting $c=e^{-\epsilon}$ and taking the limit $\epsilon\rightarrow 0$:
\begin{align}
I^{2}_c(\nu)&=c\beta_0^{-\gamma'}\int_{\log\frac{1}{c}}^\infty\left(\frac{1}{t}\left(1-\frac{\beta_1}{\beta_0^2}\frac{\log t}{t}\right)\right)^{\gamma'}\frac{dt}{c+\nu e^{-t}}\\
\leq & \beta_0^{-\gamma'}\int_{\epsilon}^\infty\left(\frac{1}{t}\left(1-\frac{\beta_1}{\beta_0^2}\frac{\log t}{t}\right)\right)^{\gamma'}\frac{dt}{1+\frac{\nu}{e^{-\epsilon}} e^{-t}}=I^{2}_{1-\epsilon}(\nu)
\end{align}
but the upper bound is in fact asymptotic since varying $c$  is equivalent to a change of scheme.
We now expand $I^{2}_{1-\epsilon}(\nu)$:
\begin{equation}
I^{2}_{1-\epsilon}(\nu)\sim \beta_0^{-\gamma'}\int_{\epsilon}^{\infty}\frac{1}{t^{\gamma'}}\left(1-\frac{\beta_1\gamma'}{\beta_0^2}\frac{\log t}{t}\right)\frac{dt}{1+\frac{\nu}{e^{-\epsilon}}e^{-t} }
\end{equation}
The first term is equal to $I^1_{1-\epsilon}(\nu)$, while the second one is the new contribution.
This new term can be linked again to the polylogarithmic function using the relation:
\begin{equation}
t^{-\gamma'-1}\log t=-\frac{\partial}{\partial \alpha} t^{-\alpha}\biggl|_{\alpha=\gamma'+1}
\end{equation} 
We find :
\begin{equation}
I_{1-\epsilon}^{2}(\nu,-\gamma')\sim I_{1-\epsilon}^1(\nu,-\gamma')+\frac{\beta_1\gamma'}{\beta_0^2}\frac{\partial}{\partial\alpha}I^1_{1-\epsilon}(\nu,-\alpha)\biggl|_{\alpha=\gamma'+1}
\end{equation}
We take the limit $\epsilon\rightarrow 0$ and we perform the derivative in the asymptotic expression of $I_1^1(\nu,-\alpha)$ in Eq.(\ref{eqn:esp_polylog}). Keeping only the leading contribution we obtain:
\begin{align}
\frac{\partial}{\partial\alpha}I_1^1(\nu,-\alpha)\biggl|_{\alpha=\gamma'+1}&=
\beta_0^{-\gamma'}\frac{\Gamma(-\gamma')}{\Gamma(-\gamma'+1)}(\log \nu)^{-\gamma'}\log\log \nu=\\
&=-\frac{\beta_0^{-\gamma'}}{\gamma'}(\log \nu)^{-\gamma'}\log\log \nu
\end{align}
Thus the asymptotic behavior to the next-to-leading logarithmic order is:
\begin{align}
I^{2}_c(\nu)&\sim  \frac{[\beta_0\log(\frac{p^2}{\Lambda_{\overline{W}}^2})]^{-\frac{\gamma_0}{\beta_0}L+1}}
{\gamma_0 L-\beta_0}
-\frac{\beta_1}{\beta_0^2}(\beta_0\log\frac{p^2}{\Lambda_{\overline{W}}^2})^{-\frac{\gamma_0}{\beta_0}L}\log\log\frac{p^2}{\Lambda_{\overline{W}}^2}\nonumber\\
& \sim \frac{1}{\gamma_0 L-\beta_0}\Biggl[\frac{1}{\beta_0\logplw}\biggl(1-\frac{\beta_1}{\beta_0^2}\frac{\log\logplw}{\logplw}\biggr)\Biggr]^{\frac{\gamma_0}{\beta_0}L-1}
\end{align}
that agrees perfectly with the $RG$ estimate Eq.(\ref{eqn:esp_ntl}).

\thispagestyle{empty}
\appendix 

\section{Fourier Transforms}\thispagestyle{empty}

In this appendix we compute Fourier transforms of the kind: 
\begin{equation}\label{eqn:f.t.}
\int {(p^2)}^L\biggl(\log\frac{p^2}{\mu^2}\biggr)^n e^{ip \cdot x}  \frac{d^4p}{(2\pi)^4} 
\end{equation}
for $L$ and $n$ positive integers.
We start writing:
\begin{equation}
{(p^2)}^L\biggl(\log\frac{p^2}{\mu^2}\biggr)^n =
{(\mu^2)}^L\frac{\partial^n}{\partial \alpha^n}\left(\frac{p^2}{\mu^2}\right)^{\alpha}\biggl |_{\alpha=L}
\end{equation}
and:
\begin{equation}
\left(\frac{p^2}{\mu^2}\right)^{\alpha}=\frac{1}{\Gamma(-\alpha)}
\int_0^{\infty}e^{-\frac{p^2}{\mu^2}t}t^{-\alpha-1}dt
\end{equation}
Substituting in Eq.(\ref{eqn:f.t.}) and exchanging the order of integration we get:
\begin{equation}
\frac{{(\mu^2)}^L}{(2\pi)^4}\frac{\partial^n}{\partial\alpha^n}
\left(\frac{1}{\Gamma (-\alpha)}\int_0^{\infty} t^{-\alpha-1}\int e^{-\frac{p^2}{\mu^2}t+ip_\alpha x_\alpha}d^4p\, dt\right)\biggl|_{\alpha=L}
\end{equation}
The integral on $p$ is now Gaussian and we obtain:
\begin{align}
&\frac{{(\mu^2)}^L}{(2\pi)^4}\frac{\partial^n}{\partial\alpha^n}
\left(\frac{1}{\Gamma (-\alpha)}\int_0^{\infty} t^{-\alpha-1}\frac{\pi^2\mu^4}{t^2}e^{-\frac{1}{4}\frac{x^2\mu^2}{t}}dt\right)\biggl|_{\alpha=L}\\
&=\frac{\pi^2}{(2\pi)^4}{(\mu^2)}^L\mu^4\frac{\partial^n}{\partial\alpha^n}
\left(\frac{1}{\Gamma (-\alpha)}\int_0^{\infty} t^{-\alpha-3}e^{-\frac{1}{4}\frac{x^2\mu^2}{t^2}}dt\right)\biggl|_{\alpha=L}
\end{align}
We compute the last integral reducing it to a $\Gamma$ function by the substitution $t'=\frac{1}{4}\frac{x^2\mu^2}{t}$:
\begin{align}
&\int_0^{\infty} t^{-\alpha-3}e^{-\frac{1}{4}\frac{x^2\mu^2}{t}}dt\\
&=\int_0^{\infty} e^{-t'} \left[\frac{x^2\mu^2}{4}(t')^{-1}\right]^{-\alpha-3}\frac{x^2\mu^2}{4}{(t')}^{-2}dt'\\
&=\left(\frac{x^2\mu^2}{4}\right)^{-\alpha-2}\Gamma\left(\alpha+2\right)
\end{align} 
The Fourier transform now reads:
\begin{align}\label{eqn:app_tf_intermedia}
&\int {(p^2)}^L\biggl(\log\frac{p^2}{\mu^2}\biggr)^n e^{ip \cdot x}  \frac{d^4p}{(2\pi)^4} \nonumber\\
&=\frac{{(\mu^2)}^{L+2}}{4(2\pi)^2}\frac{\partial^n}{\partial\alpha^n}
\left(\frac{1}{\Gamma (-\alpha)}\left(\frac{x^2\mu^2}{4}\right)^{-\alpha-2}\Gamma\left(\alpha +2\right)\right)\biggl|_{\alpha=L}
\end{align}
We evaluate the Fourier transform in some cases by means of Mathematica.
In particular we are interested in the cases $L=2$ with $n=1,2,3$.
We obtain:
\begin{align}
\int {(p^2)}^2\log\frac{p^2}{\mu^2} e^{ip \cdot x}  \frac{d^4p}{(2\pi)^4} &=
-\frac{2^6\cdot 3}{\pi^2 x^8}\nonumber\\
\int {(p^2)}^2\biggl(\log\frac{p^2}{\mu^2}\biggr)^2 e^{ip \cdot x}  \frac{d^4p}{(2\pi)^4} &=
\frac{2^7\cdot 3}{\pi^2 x^8}\bigl(-\frac{10}{3}+2\gamma_E -\log\frac{4}{x^2\mu^2}\bigr)\nonumber\\
\int {(p^2)}^2\biggl(\log\frac{p^2}{\mu^2}\biggr)^3 e^{ip \cdot x}  \frac{d^4p}{(2\pi)^4} &=
\frac{2^6\cdot 3}{\pi^2 x^8}\bigl(-\frac{51}{2}+40\gamma_E-12\gamma_E^2+\nonumber\\
&-(20-12\gamma_E)\log\frac{4}{x^2\mu^2} -3\log^2\frac{4}{x^2\mu^2}\bigr)\nonumber\\
\end{align}
where $\gamma_E$ is Euler-Mascheroni constant. \par
Moreover, again by means of Mathematica, we evaluate Eq.(\ref{eqn:app_tf_intermedia}) for generic $L$ and $n=1$ or $n=2$:
\begin{align}\label{eqn:app_tf_L_1}
&\int {(p^2)}^L\biggl(\log\frac{p^2}{\mu^2}\biggr) e^{ip \cdot x}  \frac{d^4p}{(2\pi)^4}=
-\frac{(-4)^L  L!  \Gamma(2+L)}{\pi ^2}x^{-2 (2+L)}\\
&\int {(p^2)}^L\biggl(\log\frac{p^2}{\mu^2}\biggr)^2 e^{ip \cdot x}  \frac{d^4p}{(2\pi)^4}\nonumber\\ 
\label{eqn:app_tf_L_2}
&=2\frac{(-4)^L  L! \Gamma(2+L) 
}{\pi ^2}\left(\gamma_E-H(L)-\psi(2+L)+\log\left(\frac{x^2 \mu ^2}{4}\right)\right)x^{-2 (2+L)}
\end{align}
where $H(L)$ is the harmonic number defined by:
\begin{equation}
H(L)=\sum_{i=1}^{L}\frac{1}{i}
\end{equation}
and $\psi$ is the digamma function defined by:
\begin{equation}
\psi(z)=\frac{\Gamma'(z)}{\Gamma(z)}
\end{equation}
Inverting Eqs.(\ref{eqn:app_tf_L_1}-\ref{eqn:app_tf_L_2}) we obtain:
\begin{align}\label{eqn_app_tf_inv_1}
&\int x^{-2 (2+L)} e^{-ip\cdot x}d^4x =
 - \frac{\pi^2}{(-4)^L L! \Gamma(2+L)}{(p^2)}^L\biggl(\log\frac{p^2}{\mu^2}\biggr)\\\label{eqn_app_tf_inv_2}
&\int x^{-2 (2+L)} \log\left(\frac{x^2 \mu ^2}{4}\right)e^{-ip\cdot x}d^4x =
\frac{1}{2}\frac{\pi^2}{(-4)^L L! \Gamma(2+L)}{(p^2)}^L\log^2\frac{p^2}{\mu^2}+\nonumber\\
&+\frac{\pi^2}{(-4)^L L! \Gamma(2+L)}\left(\gamma_E-H(L)-\psi(2+L)\right){(p^2)}^L\log\frac{p^2}{\mu^2}
\end{align}

We are also interested in extracting the leading logarithms in Eq.(\ref{eqn:app_tf_intermedia}) in the generic case.
We obtain the leading logarithm from the terms that contain $n-1$ derivatives with respect to $\alpha$ of $\left(\frac{x^2\mu^2}{4}\right)^{-\alpha-2} $ and one derivative of $\frac{1}{\Gamma(-\alpha)}$, since otherwise we get zero because $\frac{1}{\Gamma(-L)}=0$ for $L$ a positive integer: 
\begin{align}
&\int {(p^2)}^L\biggl(\log\frac{p^2}{\mu^2}\biggr)^n e^{ip \cdot x}  \frac{d^4p}{(2\pi)^4} \nonumber\\
&=\frac{n\Gamma(L+2) 2^{2L}}{\pi^2}\biggl(\frac{\Gamma'(-\alpha)}{\Gamma^2(-\alpha)}\biggr)\biggl|_{\alpha\rightarrow L}
\frac{1}{(x^2)^{L+2}}\biggl(\log\frac{4}{x^2\mu^2}\biggr)^{n-1}+\cdots
\end{align}
The factor of $n$ occurs because there are $n$ such terms performing the $n$-th derivative.

The limit $\biggl(\frac{\Gamma'(-\alpha)}{\Gamma^2(-\alpha)}\biggr)\biggl|_{\alpha\rightarrow L}$ can be easily calculated knowing that the residue of the gamma function at $-L$
is $\frac{(-1)^L}{L!}$. The result is:
\begin{equation}
\biggl(\frac{\Gamma'(-\alpha)}{\Gamma^2(-\alpha)}\biggr)\biggl|_{\alpha\rightarrow L}=
(-1)^{L+1}L!
\end{equation}
Therefore, the leading logarithm of the Fourier transform is:
\begin{align}\label{eqn:ft_leading}
&\int {(p^2)}^L\biggl(\log\frac{p^2}{\mu^2}\biggr)^n e^{ip \cdot x}  \frac{d^4p}{(2\pi)^4} \nonumber\\
&=\frac{n\Gamma(L+2) 2^{2L}}{\pi^2}(-1)^{L+1} L!
\frac{1}{(x^2)^{L+2}}\biggl(\log\frac{4}{x^2\mu^2}\biggr)^{n-1}+\cdots
\end{align}

\thispagestyle{empty}

\end{document}